\documentclass[twocolumn,secnumarabic,amssymb, nobibnotes, aps, prx,superscriptaddress]{revtex4-1}

\usepackage[paperwidth=210mm,paperheight=297mm,centering,hmargin=2cm,vmargin=2.5cm]{geometry}
\usepackage{amsmath} 
\usepackage{graphicx} 
\usepackage{listings} 
\usepackage{soul}
\usepackage{color}
\usepackage{lipsum}
\usepackage{float}
\usepackage{multirow}
\usepackage{adjustbox}
\usepackage{booktabs}

\begin{document}

\title{Deciphering Molecular Charge Anisotropy:\\the Case of Antibody Solutions}

\author{Fabrizio Camerin}
\email[Corresponding author:]{ fabrizio.camerin@chem.lu.se}
\affiliation{Division of Physical Chemistry, Department of Chemistry, Lund University, Lund, Sweden}

\author{Susana Marín-Aguilar}
\affiliation{Department of Physics, Sapienza University of Rome, Piazzale Aldo Moro 2, 00185 Roma, Italy}

\author{Anna Stradner}
\affiliation{Division of Physical Chemistry, Department of Chemistry, Lund University, Lund, Sweden}
\affiliation{LINXS Institute of Advanced Neutron and X-ray Science, Lund University, Lund, Sweden}

\author{Peter Schurtenberger}
\email[Corresponding author:]{ peter.schurtenberger@chem.lu.se}
\affiliation{Division of Physical Chemistry, Department of Chemistry, Lund University, Lund, Sweden}
\affiliation{LINXS Institute of Advanced Neutron and X-ray Science, Lund University, Lund, Sweden}

\author{Emanuela Zaccarelli}
\email[Corresponding author:]{ emanuela.zaccarelli@cnr.it}
\affiliation{CNR Institute of Complex Systems, Uos Sapienza, Piazzale Aldo Moro 2, 00185 Roma, Italy}
\affiliation{Department of Physics, Sapienza University of Rome, Piazzale Aldo Moro 2, 00185 Roma, Italy}

\date{\today}

\begin{abstract}
\noindent

Electrostatic interactions fundamentally govern the structure, stability, and dynamics of charged (bio)matter, yet the impact of heterogeneous and anisotropic charge distributions on the behavior of protein solutions remains elusive. Here, we introduce a versatile multiscale framework that directly connects molecular-level electrostatics to collective properties via a colloid-inspired coarse-grained modeling combined with neural network-assisted optimization. Using monoclonal antibodies as model system, our inverse design approach identifies charge patterns capable of reliably reproducing experimental structure factors, osmotic compressibility and collective diffusion coefficients in a wide region of protein concentrations. 
Close inspection of our data further uncovers how specific physical features and spatial arrangements of localized charge patches significantly influence the solution structure. This transferable strategy provides a predictive pathway to decode and control charge-driven interactions in complex biomolecules and, more generally, in heterogeneously-charged soft matter systems, with immediate relevance to protein formulation and biomaterials engineering.

\end{abstract}

\maketitle

Among the manifold features governing microscopic systems, electrostatic interactions are a key force shaping the behavior of matter that surrounds us.
From colloidal suspensions and polyelectrolytes to proteins and DNA, the presence and spatial distribution of electric charges dictate how individual components interact, self-organize and evolve in time~\cite{wong2010electrostatics,rubinstein2012polyelectrolytes,ruiz2021role}. Additionally, external fields can be used to influence and induce specific behaviors of particles and assemblies that are endowed with charges~\cite{nojd2013electric,vutukuri2014experimental,colla2018self}. In soft matter systems, electrostatics can stabilize crystalline or dispersed phases~\cite{zang2024enabling}, promote phase separation~\cite{notarmuzi2024features} or, for instance, mediate the formation of complex supramolecular architectures~\cite{dean2014electrostatics,hueckel2021total}. In biological environments, these interactions acquire even greater significance due to the chemical diversity and structural complexity of biomolecules~\cite{bianchi2020relevance,ren2012biomolecular,zhou2018electrostatic}.
Biological systems constitute a fertile playground for investigating electrostatic effects, since they often carry heterogeneous and anisotropic charge distributions that have been finely tuned by evolution to perform specific functions~\cite{warshel2006electrostatic,zhou2018electrostatic}. 

Monoclonal antibodies are a particularly compelling example, owing both to their widespread therapeutic relevance and to their suitability as model systems for studying collective phenomena in soft matter and biophysics~\cite{pantaleo2022antibodies,singh2018monoclonal,castelli2019pharmacology,skar2019colloid,skar2023using}.
Antibodies are complex proteins that are anisotropic in shape and often also in charge distribution. Therefore, they can be considered as anisotropic colloids, whose biological functionality and physicochemical behaviors are intimately connected through their specific amino acid sequences. Their  primary protein structure not only defines antigen-recognition sites, but also establishes distinct spatial patterns of charged and hydrophobic residues across the molecular surface~\cite{tsumoto2019future,chiu2019antibody}. Collectively, these characteristics dictate interactions with solvent, ions, and other molecules in solution. Consequently, environmental conditions such as pH and ionic strength critically shape antibody behavior by altering amino acid protonation states and modulating electrostatic screening, ultimately impacting the net charge distribution and intermolecular interaction landscape~\cite{camerin2025electrostatics,polimeni2024multi}.
The complexity introduced by this interplay between sequence-driven physical chemistry and solution-dependent interactions confers antibodies with remarkable specificity and adaptability, while also posing substantial challenges in their formulation. Achieving a comprehensive understanding of how distinct electrostatic profiles influence antibodies intermolecular interactions, as well as the static and dynamic properties of their solutions, is thus essential. Ultimately, such insights will enable precise control over their formulation, optimizing  stability, efficacy, and suitability for therapeutic administration to patients~\cite{davis2024subcutaneous,filipe2012immunogenicity,roberts2014protein,pham2020protein}. 
While the electrostatic landscape of antibodies is encoded at the level of individual amino acids, the collective properties of interest -- such as structure factors, compressibility, and transport -- are accessed experimentally only through ensemble-averaged measurements on many interacting molecules.
As a result, experiments alone cannot disentangle how specific spatial features of charge heterogeneity give rise to the observed macroscopic behavior, nor can atomistic simulations directly access the relevant concentration and length scales.
A physically grounded model is therefore required to bridge molecular-scale electrostatics and experimentally measured collective properties.

To construct such a link, it is thus crucial to simplify the complexity of the problem and adopt an alternative perspective with respect to the full atomistic treatment~\cite{polimeni2024multi,hatch2024anisotropic,vinterbladh2025intermolecular,pineda2025patchy}. For the specific case of electrostatic interactions, theoretical and computational approaches, such as those based on the DLVO theory~\cite{derjaguin1993theory,hansen2000effective,gnidovec2025anisotropic}, have been proposed to incorporate charge effects and their screening through effective potentials. However, in many cases, their applicability is often limited both for predicting static solution properties and for interpreting collective dynamic phenomena or many-body effects~\cite{trefalt2017forces,camerin2025electrostatics,ter2025machine}. 
More sophisticated methods have attempted to explicitly account for multipolar contributions or detailed spatial charge distributions, as exemplified by Poisson–Boltzmann or generalized Born theories~\cite{fogolari2002poisson,baker2005biomolecular,onufriev2019generalized}. However, despite their enhanced complexity and computational expense, these advanced approaches have yet to be employed for a comprehensive understanding of collective behavior of macromolecules in solution.

Conversely, coarse-grained modeling approaches, inspired by established strategies from soft matter and colloidal science, allow one to reduce unnecessary molecular-level details and focus instead on essential physical interactions that dominate antibody behavior~\cite{stradner2020potential,skar2019colloid}. In these simplified representations, single beads typically represent clusters of atoms or even amino acids, and their connectivity is designed to reproduce the overall shape of the molecules. This reduction in complexity makes ensemble-level properties directly accessible in simulations. In fact, using coarse-grained bead-based models featuring explicit Coulomb charges, some of us  have recently been able to capture qualitative trends in the viscosity of antibody solutions~\cite{camerin2025electrostatics,johnston2025viscosity}. This was shown to result from the formation of correlated structures arising from the long-range nature of the electrostatic interactions, which did not occur for simple screened Coulomb effective potentials.
While this modeling approach was performed on an antibody characterized by a large net positive charge and deprived of \textit{evident} patchy regions, a natural question arises on the extent by which an explicit charge model can capture the behavior of antibodies (or other biological molecules) with markedly heterogeneous charge distributions. 
Interestingly, for heterogeneously charged antibodies, it has been shown that the use of screened Coulomb interactions is even inadequate to describe their solution structure and thermodynamics, while with an explicit treatment of charges the correct behavior has been recovered~\cite{camerin2026beyond}.
These findings were obtained for a weakly coarse-grained model where each amino acid was represented by a bead, which is not suitable for systematic studies of collective dynamic properties through molecular dynamics simulations. 
It is therefore essential to introduce a further level of coarse-graining based on an effective and robust mapping strategy, in which the heterogeneous charge distribution of the molecular model is transferred to the coarse-grained representation while preserving the electrostatic features that govern solution properties. Currently, coarse-grained parametrization of anisotropic and structurally complex molecules largely rely on trial-and-error procedures that are difficult to generalize and rarely incorporate charge information directly inherited from detailed molecular descriptions~\cite{chowdhury2020coarse,samanta2018influence}. In this context, being able to quantify and model the spatial arrangement of charges is valuable \textit{per se}, as it can both improve predictive power and provide mechanistic insight into solution behavior. 

In light of these considerations, here we establish a comprehensive and broadly applicable framework for elucidating the role of distinct charge distributions in antibodies, as a prototypical case of anisotropic (bio)molecules. 
We systematically investigate how variations in internal charge distributions impact the solution structure, demonstrating that even subtle changes can significantly alter its behavior, highlighting the importance of detailed electrostatic characterization. Furthermore, we introduce an approach bridging micro and mesoscale molecular features to build physically meaningful models capable of accurately predicting both static and dynamic properties.
Central to our strategy is a conceptually straightforward and computationally efficient machine learning (ML) methodology that is seamlessly combined with traditional physical modeling and molecular dynamics (MD) simulations. This approach, which represents an innovative addition to existing methodologies in biophysics~\cite{mamoshina2016applications,kim2023computational}, is able to efficiently predict optimal coarse-grained charge distributions, providing a robust and practical solution for capturing electrostatic anisotropy with minimal computational overhead. By leveraging comparison with experimental data from small-angle X-ray scattering, through an \textit{inverse-design} approach, we identify and reconstruct the most accurate charge representations under relevant experimental conditions. 
Finally, by exploiting liquid-state theory, we connect microscopic electrostatic features to experimentally accessible collective properties of antibody solutions.
Our study not only enhances the accuracy and interpretability of antibody modeling but also provides a versatile blueprint applicable to a wide array of electrostatically governed phenomena in anisotropic soft matter and biological systems.

\section*{Results and Discussion}

\subsection*{Choice of the model architecture}

The first step for analyzing the effect of different charge distributions is the creation of a sound model, which retains the structural peculiarities of the molecule and allows for correctly reproducing the salient features of its internal charge distribution that determine the protein-protein interactions. Despite the target being a coarse-grained model through which we can analyze ensemble properties, it is actually more convenient to start from a more finely resolved model, such as the amino acid representation of the antibody. On the one hand, it gives the possibility to calculate the (approximate) total charge retained by the molecule under specific experimental conditions; on the other hand, we can gain information on the protonation state of different amino acids, and thus the overall charge distribution within its structure. 

\begin{figure*}[t!]
\centering
\includegraphics[width=\textwidth]{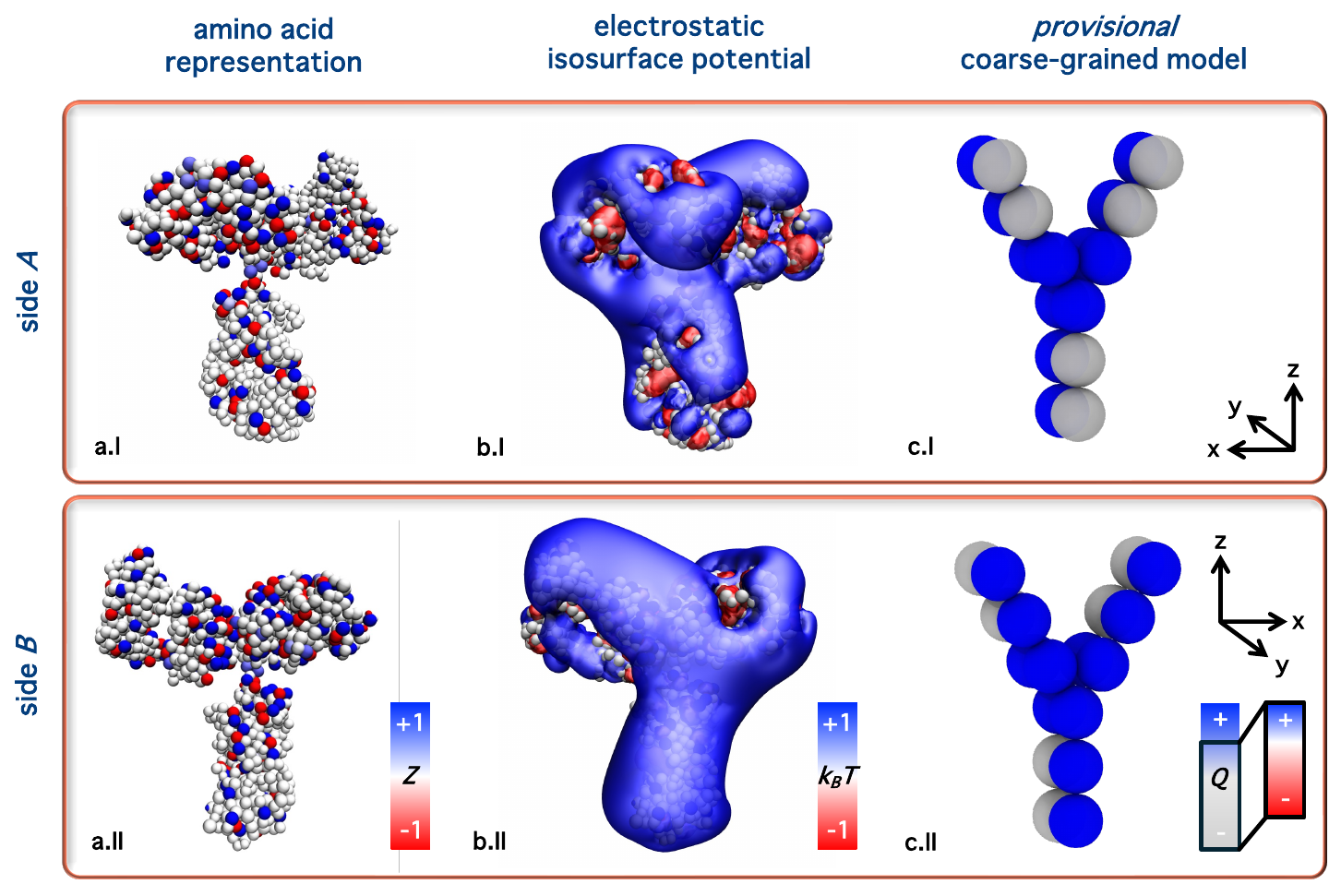}
\caption{\small \textbf{Modeling strategy.} (a) Amino acid representation of the antibody under investigation. The color coding reflects the charge of each amino acid. (b) Electrostatic isopotential surfaces for the same antibody as in (a). The color coding reflect the potential of the surface at $\pm 1k_BT/e$. (c) Provisional 18-bead coarse-grained model for the antibody in which two 9-bead layers are over-imposed onto each other. Blue beads show the (likely) charge assignment given the representation in (b), while grey beads remain undetermined given the complex arrangement of charges in the corresponding regions. In all cases, panels I and II provide the visualization of the same antibody from two different perspectives, highlighting the two sides of the antibody, side $A$ and $B$, respectively, as indicated by the reported axes.}
\label{fig:provisionalmodel}
\end{figure*}

Here, we investigate a recombinant chimeric human/mouse IgG1 monoclonal antibody (see Methods), for which an atomistic-to-amino acid coarse-grained representation and an experimental characterization was recently obtained by some of us~\cite{camerin2026beyond}. This specific antibody was calculated to have an overall charge $Q \approx +24$ at a $pH=6$ and ionic strength $I=7$ mM, and was shown to be colloidally stable without any arrested states or phase separation up to high concentrations $c \lesssim 200$ mg/ml under these conditions.
The amino acid representation of the antibody is reported in Figure~\ref{fig:provisionalmodel} from two different perspectives exposing sides $A$ and $B$, respectively. The color coding indicates the average charge of each individual amino acid as defined by the color bar also reported in the Figure. We visualize the charge distribution on the antibody structure by showing in panel (b.I-II) the corresponding electrostatic isopotential surfaces at $\pm 1 k_BT/e$, obtained with an \textsc{apbs} solver (see Methods). It appears evident that the antibody under investigation is endowed, for a large part of its structure, with a positive charge. Nonetheless, side $A$ also displays regions where the underlying amino acids give rise to neutral and even negative areas, as evidenced by the red surfaces embracing the antibody close to the most external regions of its domains.

Based on the successful modeling obtained in previous works~\cite{polimeni2024multi,camerin2025electrostatics}, we thus build a coarse-grained model inspired by colloidal physics in which the structure of the antibody is made up by spheres at contact forming a Y-shaped scaffold. For the cases investigated previously, the typical structure of the antibody could be reasonably reproduced using 9 beads. However, for the present case, due to the peculiar charge distribution of the antibody, it appears more reasonable to design a model based on two layers which, for the sake of simplicity, are placed at a distance $\sigma$ with respect to each other, with $\sigma$ being the size of each bead. The resulting 18-bead architecture of the coarse-grained model is shown in Figure~\ref{fig:provisionalmodel}(c). 
In order to assign charges to each bead, in Ref.~\cite{camerin2025electrostatics} the model and the amino acid representation were made to overlap, and the individual amino acid charges were then transferred to the closest bead of the coarse-grained structure. There,  the examined antibody had an almost uniform charge distribution, thereby (likely) facilitating the mapping. In the current case, however, the charge distribution appears to be more fragmented on the antibody surface, with negatively charged areas quite concentrated in specific regions on side $A$, which likely could act as attractive patches. The complexity of the charge distribution makes a direct mapping from the amino acid model much less straightforward, as small changes in the rescaling of the overall size performed for the overlap between the two structures would give rise to rather different charge distributions on the coarse-grained model. Still, based on the electrostatic isopotential surface, it seems reasonable to assume that one of the two layers (side $B$) is quite uniformly positively charged, while on side $A$ the positive charge is concentrated in the central region of the model (see Figure~\ref{fig:provisionalmodel}c). It remains however an open question how to properly assign these charges on side $A$ and how the precise location of the negatively charged attractive sites influences the overall solution behavior. These aspects will be investigated in the following.

\subsection*{Exploring the consequences of a variation of the charge distribution}

\begin{figure*}[t!]
\centering
\includegraphics[width=\textwidth]{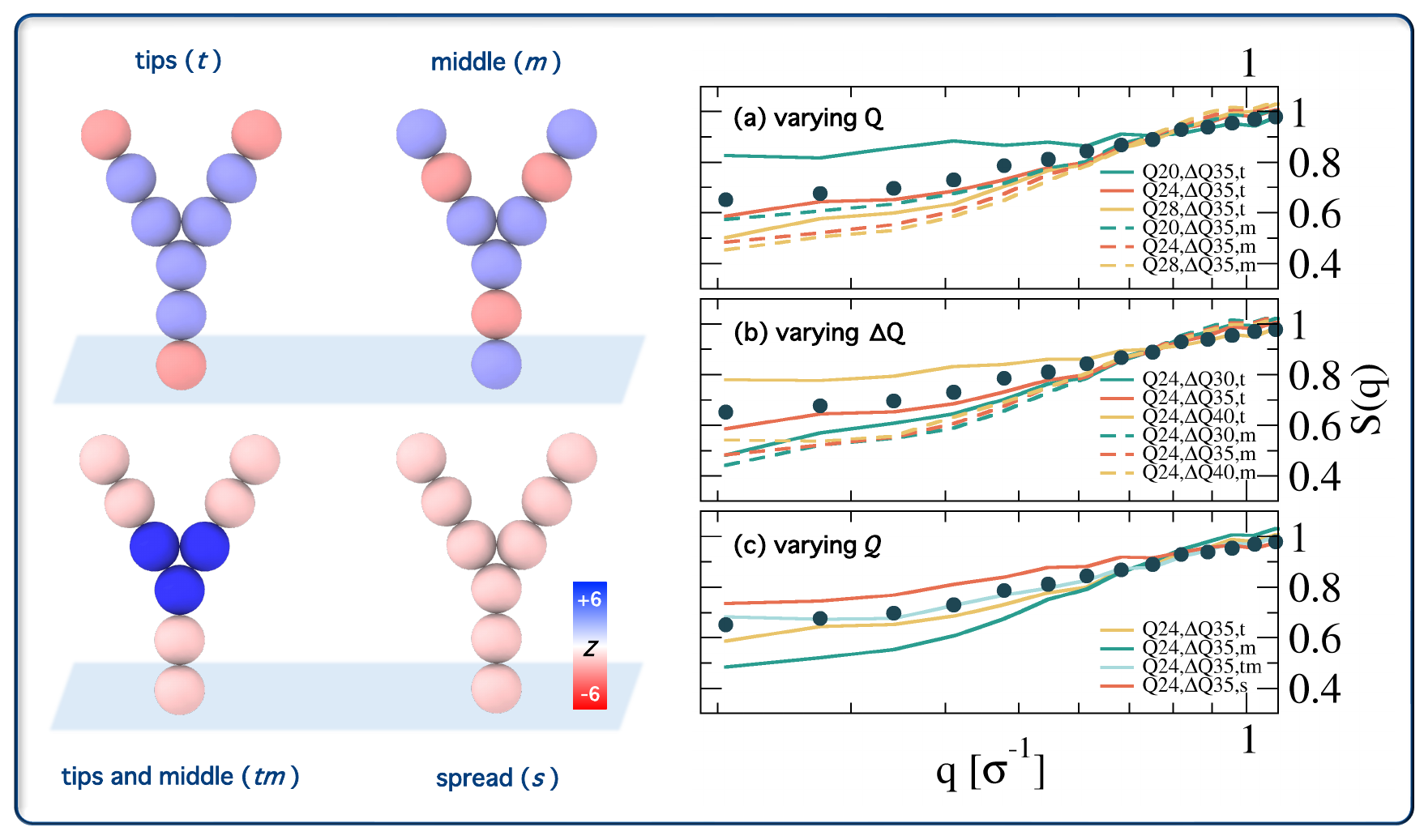}
\caption{\small \textbf{Varying parameters.} (Left) Tips $t$, middle $m$, tips and middle $tm$, and spread $s$ representative charge distributions on side $A$ of the antibody model that vary depending on the position of the negative charges (red beads). Side $B$ is kept fixed in all cases and it is not shown in the Figure. The color coding refers to the charge $Z$ assigned to each bead. (Right) Static structure factor $S(q)$ as a function of the scattering vector $q$ in units of $\sigma^{-1}$, with $\sigma$ the bead size, for (a) varying the total charge $Q$, (b) varying the charge difference $\Delta Q$ and (c) the charge distribution $\mathcal{Q}$, keeping fixed the other parameters, for $c=20$ mg/ml. Indicated in the legends is the nominal value of the parameters, while the actual one may slightly change (see Results and Methods). As a reference, circles indicate the experimental $S(q)$ for $c=20$ mg/ml, taken from Ref.~\cite{camerin2026beyond}.}
\label{fig:varyingparameters}
\end{figure*}

Before attempting to develop a protocol for assigning the charges $\tilde{q_i}$ to each bead $i$ in the coarse-grained model, we assess the consequences of varying the charge distribution on side $A$ of the 18-bead model for the static structure factor $S(q)$, a key experimental quantity obtained by small-angle X-ray scattering (SAXS). 
We note that the explored parameter space is based on a preliminary characterization of the antibody (see Methods), yet the analysis is deliberately carried out beyond the specific constraints of the system. This allows us to identify general trends associated with variations in total charge, bead charge, and charge location that are not tied to a single, specific antibody.
To this aim, we create a wide set of configurations in which we vary (i) the total (net) charge of the antibody $Q=\sum_i^{18} \tilde{q_i} = 20,24,28$, (ii) the difference between the sum of positive and negative charges, that is the total number of positive and negative charges, $\Delta Q= Q_+-Q_- = \sum_{i: \tilde{q_i}>0} \tilde{q_i} - \sum_{i: \tilde{q_i}<0} \tilde{q_i} = \sum_i^{18} |\tilde{q_{i}}|= 30,35,40$, and (iii) the charge distribution on the structure of the antibody $\mathcal{Q}$. For the latter, we vary $\mathcal{Q}$ on side $A$ only, since, based on the electrostatic isopotential surface, side $B$ is likely endowed with positive charges arranged in (quasi-)uniform fashion. We thus create configurations in which negative beads are placed i) at the tips $\mathcal{Q}_t$, ii) in the middle $\mathcal{Q}_m$, iii) in the tips and middle $\mathcal{Q}_{tm}$ and iv) spread on all side $A$ beads $\mathcal{Q}_s$, as illustrated in Figure~\ref{fig:varyingparameters}. 
In cases where it is not possible to generate a configuration with an exact combination of the target values of $Q$, $\Delta Q$ and $\mathcal{Q}$, the algorithm produces the closest admissible configuration by minimizing the deviation from the desired parameters. This controlled approximation still allows systematic trends to be identified when varying a single parameter. Additional details of the algorithm are provided in Methods.

We then run equilibrium molecular dynamics (MD) simulations of an ensemble of antibodies at two different concentrations, specifically $c=20$ and $100$ mg/ml, at an ionic strength $I=7$ mM, which we subsequently compare to experiments. These two concentrations were selected because of their sensitivity to different contributions from protein-protein interactions. For more dilute solutions, electrostatic interactions dominate over short-range contributions such as van der Waals and hydrophobic forces, allowing for a proper assessment of the effects arising from the internal charge arrangement. Conversely, at $c = 100$ mg/mL, the solution becomes sufficiently crowded that excluded-volume interactions prevail and electrostatic contributions play a more marginal role. Based on our recent works~\cite{camerin2025electrostatics,camerin2026beyond}, we model charges with explicit Coulomb interactions. Extensive details on the MD simulations are given in Methods.

We thus calculate the static structure factor $S(q)$ of the antibody solutions, as defined in Eq.~\ref{eq:sq}, for all simulated systems. These are reported in Figure~\ref{fig:varyingparameters} for $c=20$ mg/ml for selected combinations of parameters and by varying $Q$, $\Delta Q$ and $\mathcal{Q}$. In each panel, the legend reports the nominal values which, as explained before, might slightly differ from the exact actual values. 
Furthermore, for simplicity, here we show the results obtained from configurations in which all positive and negative charges have the same values on different beads (same coloring in  Figure~\ref{fig:varyingparameters}); outcomes obtained from combinations with randomized charges on the beads and corresponding results for $c=100$ mg/ml are reported in the SI. 

We immediately notice that for the investigated parameters the variation in $S(0)$, related to the osmotic compressibility of the suspension,  is quite pronounced, reaching up to a factor of $2$, from $\approx 0.4$ to $\approx 0.8$. We can also observe a few general trends. In particular, a decrease in $Q$ at constant $\Delta Q$ results in a higher $S(0)$ (Figure~\ref{fig:varyingparameters}(a)), and the same effect can be observed for an increase in $\Delta Q$ at a fixed $Q$ (Figure~\ref{fig:varyingparameters}(b)), respectively. 
These findings stem from the balance between the total net charge, that contributes to the overall repulsive interactions between antibodies, and the difference between positive and negative charges, that is responsible for the electrostatic contribution to attraction.
This, combined with its structural arrangement -- which consists of a predominantly repulsive layer (side $B$) and one in which there are negative charges (side $A$) -- contributes to the formation of an overall more ``polarized" molecule, which enhances intermolecular attractions and ultimately leads to an increase in $S(0)$. 
While this effect is observed for both positions $\mathcal{Q}_t$ and $\mathcal{Q}_m$ of the negative charges, it appears to be significantly more pronounced for $\mathcal{Q}_t$, as shown  in Figure~\ref{fig:varyingparameters}(c), where the effect of a different charge distribution for fixed $Q$ and $\Delta Q$ is reported. The case that shows the lowest $S(0)$ occurs when charges are placed in the middle beads while the one with the highest compressibility is the one with charges spread on all the beads of side $A$. Once again, this trend can be interpreted in terms of the molecules' attractive capabilities, which are maximized with respect to other antibodies when the negative beads are distributed on one of the two layers -- being side $B$ fully positive -- and minimized when the charges are placed at the center of the molecule, where the attractive effect is screened by the outermost positive beads.

This analysis highlights the intricate interdependence among the various parameters that govern the electrostatic properties of these anisotropic molecules, as well as the non-negligible changes in solution structure that result from even slight adjustments of the charge parameters. 
Moreover, one must also account for the possibility of assigning different charges to beads within each antibody arm, rather than using identical positive or negative charges of the same magnitude as done so far for simplicity. This more realistic representation of the charge distribution arising from the amino acids significantly increases the number of possible charge combinations attainable  within the antibody structure.
At the same time, the resulting charge pattern should be preserved across different concentrations, as it reflects an intrinsic property of the antibody under fixed solution conditions.
The vastness of the parameter space, coupled with these physical constraints, renders the problem exceedingly difficult to be addressed using traditional trial-and-error simulations, in which charges on individual beads are adjusted arbitrarily by sequential simulations. Rather, a more systematic and robust methodology is required.

\subsection*{Machine Learning-assisted determination of the charge features}

In order to determine the charge distribution of the coarse-grained model that best matches the experimentally determined structure factors, we develop a protocol that is based on a neural network (NN). The key idea is to establish a mapping between the structure factor resulting from a given charge distribution and the charge distribution itself. In this way, following an initial training phase, it is possible to input an experimental structure factor and retrieve the bead-wise charge configuration that best reproduces the observed experimental data. The most important advantage of this data-driven procedure is that inherent relations and trends that are present between $S(q)$ and the charge of each bead $\tilde{q}_i$ are accounted automatically in the machine learning model, avoiding to ``manually" detect correlations in a wide phase space. 

\begin{figure*}[t!]
\centering
\includegraphics[width=\textwidth]{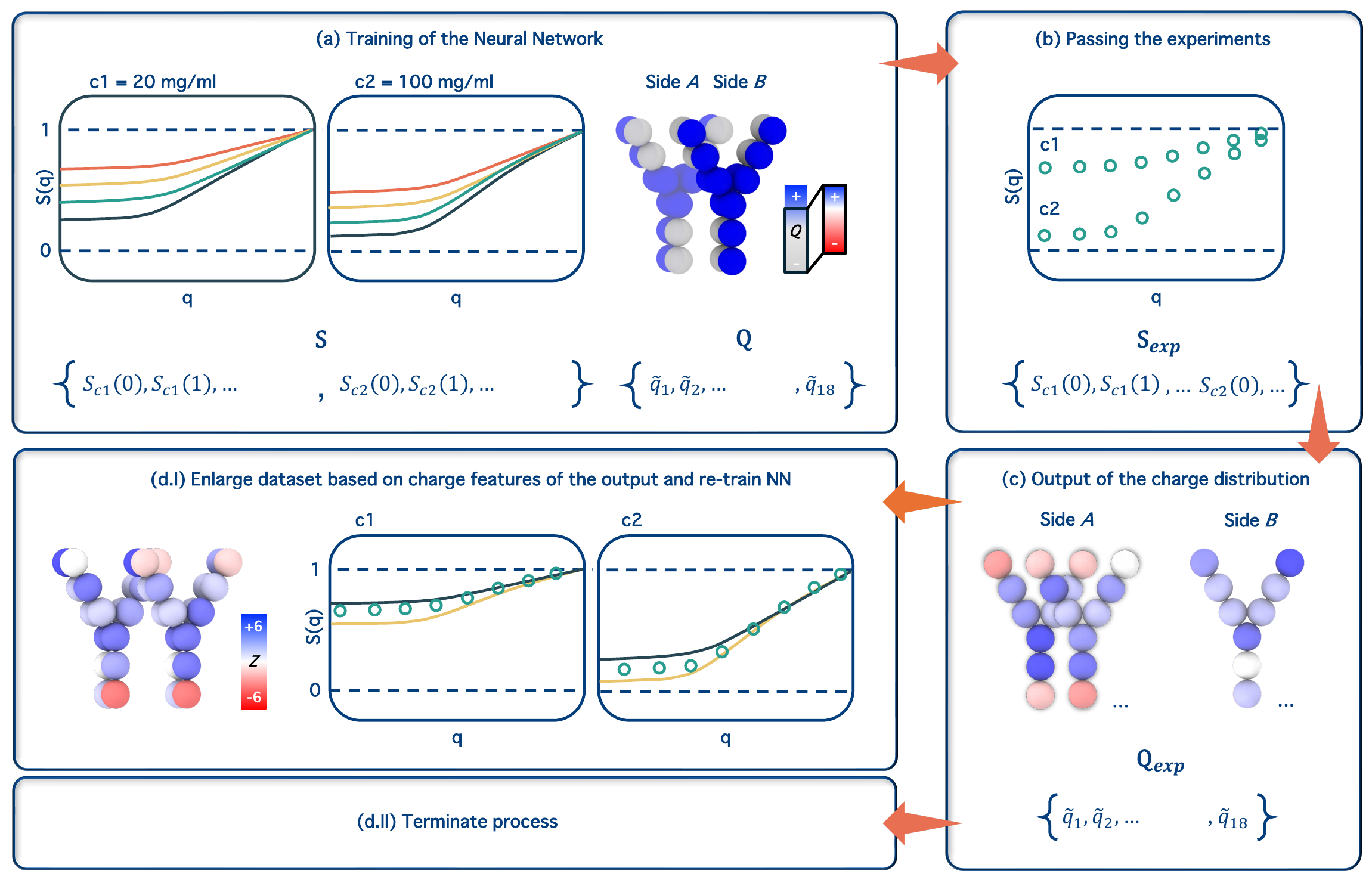}
\caption{\small \textbf{Schematics of the optimization protocol.} (a) The static structure factors $S(q)$ for two concentrations $c_1=20$ mg/ml and $c_2=100$ mg/ml obtained from a training dataset are concatenated into an array $\mathbf{S}$. Each of the rows of this array is linked to a unique charge distribution stored in another array $\mathbf{Q}$. (b) The experimental target data are arranged in an array $\mathbf{S}_{exp}$ with the same discrete scattering vectors $q$ as for $\mathbf{S}$. (c) Based on the training of the neural network performed in (a), and passing the experimental data from (b), several output charge distributions $\mathbf{Q}_{out}$ are obtained. The variability is mainly on side $A$ of the antibody since the charge distribution of side $B$ does not vary much in the training dataset. (d) Based on the physical features that are extracted from the output distributions and on the level of agreement with the target, (I) the initial training dataset can be enlarged including additional, closer-to-the-target $S(q)$s, and the overall process repeated in a new cycle, thus creating updated $\mathbf{S}$ and $\mathbf{Q}$ arrays; otherwise, (II) the protocol can be terminated.
}
\label{fig:schematicsml}
\end{figure*}

We thus begin by discussing the setup for our protocol. The training dataset for the NN is based on a number of static structure factors obtained from simulations. Given the number of parameters and the treatment of bead charges and counter- and buffer ions by means of explicit Coulomb interactions, the number of simulations and the computational time required would be extremely high for building a very detailed and extensive dataset.
For this reason, we opt to rely on optimization cycles according to which we successively provide improved datasets based on the outcomes of the previous step. In this way, we are allowed to provide a limited -- yet still fully representative -- dataset with the aim to converge within a few cycles to a solution that best represents the experimental outcomes. 
The implementation scheme is reported in Figure~\ref{fig:schematicsml}. In the first stage, we build a matrix  $\mathbf{S}$ that contains in each of the $N_{S(q)} \times N_{q}$ elements the value of the static structure factor $S(q_j)$ at the specific scattering vector $q_j$. Correspondingly, we build a second $N_{S(q)} \times N_{\tilde{q}}$
matrix $\mathbf{Q}$ that contains the charge distribution of the 18-bead model. 
Hence, for $\mathbf{S}$ and $\mathbf{Q}$, each row of the matrices is associated to a unique static structure factor or charge distribution, respectively.
Importantly, in the $\mathbf{S}$ matrix, we concatenate in the same line the structure factors for $c=20$ mg/ml and for $c=100$ mg/ml, as shown in Figure~\ref{fig:schematicsml}(a). 
The construction of a single-network with shared constraint across concentrations has the role of assigning a unique charge distribution to the static structure factors of both concentrations, as per physical requirement discussed in the previous section. Extensive details on the architecture of the NN are provided in Methods. 

Once the training phase is completed, we provide as input the experimental structure factors of both concentrations (Figure~\ref{fig:schematicsml}(b)), obtaining a number of outputs that reflect the best possible combinations of charges for side $A$ and $B$ (Figure~\ref{fig:schematicsml}(c)). For the former, we expect the highest variability in the outcomes. 
At this stage, we run the corresponding simulations and measure the level of agreement with the target experiments. Therefore, based on the physical electrostatic features that provide (one of) the best agreements, we enlarge the training dataset by running additional simulations with a charge distribution that is closer to the best outcome(s) (Figure~\ref{fig:schematicsml}(d)). This, together with the previously formed dataset, determines the new training data for the subsequent optimization cycle. We now explore how this conceptually simple protocol can be used for the specific case of the antibody under investigation.

\begin{figure*}[t!]
\centering
\includegraphics[width=\textwidth]{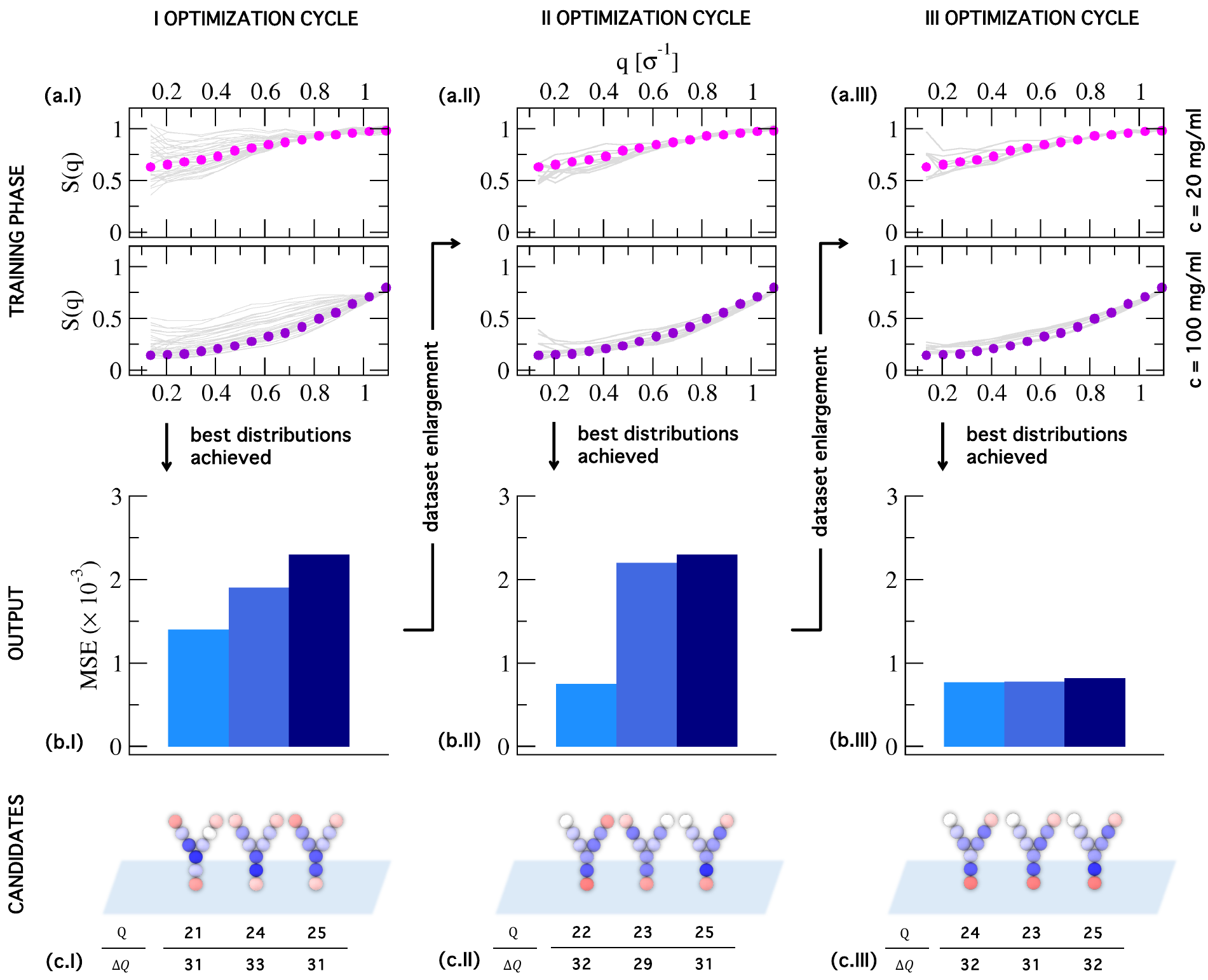}
\caption{\small \textbf{Optimization cycles.} For each of the three optimization cycles (I, II, III), we show the relevant data for (a) training phase, (b) output and (c) charge distribution candidates. In (a), we show the static structure factors $S(q)$ (grey lines) used in each cycle for the training dataset for (upper panels) $c_1=20$ mg/ml and (lower panels) for $c_2=100$ mg/ml. Symbols are for the respective experimental data, taken from Ref.~\cite{camerin2026beyond}. (b) The level of agreement of the output with experimental data is then measured for several different configurations (three are shown) by means of the mean squared error (MSE). (c) The corresponding charge distributions for side $A$ are displayed correspondingly while, for side $B$, variations are minimal. For each charge distribution, the values of $Q$ and $\Delta Q$ are also reported. Additional data are shown in the SI.}
\label{fig:optimizationml}
\end{figure*}

To begin with, we prepare the first dataset moving from the simulation results discussed in the previous section. In fact, they include different charge distributions for a range of charge values that are reasonable for the antibody of interest. Additionally, in order to introduce variability in the dataset, we randomize the charge distribution within each antibody while keeping the same $Q$, $\Delta Q$ and $\mathcal{Q}$ introduced before (see Methods). The full initial training dataset is reported in the SI. In this way, the total number of structure factors used, that is $N_{S(q)}$, is $\approx 30$. Individual $S(q)$ that were found to be too distant from the target ($\Delta S(0) \gtrsim 0.4$) were manually removed. In Figure~\ref{fig:optimizationml}(a.I) we report representative structure factors for both concentrations, with the aim of showing the wide range of $S(0)$ covered in this first data set. The full set of $S(q)$ arising from this dataset is reported in the SI. 
In Figure~\ref{fig:optimizationml}(b.I), we report the mean squared error (MSE) as compared to the experimental structure factors obtained by running simulations with few of the best outcomes of the NN in the first optimization cycle (see Methods). It is interesting to note that these already show satisfactory agreement with experiments at both concentrations, with a mean square error of $O(10^{-3})$. 
The snapshots of side $A$ reported in Figure~\ref{fig:optimizationml}(c.I) reveal that the candidates share remarkably common features. At first sight, we note that (i) negative charges are placed at the tips of the antibody while their central part is characterized by the presence of positively charged beads, being reminiscent of a $\mathcal{Q}_t$ charge distribution; we do not report side $B$ of the antibody since there are little or no deviations compared to the initial guess. 
Looking at the specific $Q$ and $\Delta Q$ reported at the bottom of Figure~\ref{fig:optimizationml}(c), it also appears that the balance between charge features is such that candidates have (ii) a medium to low overall charge, among the probed ones, that favors an increase in $S(q)$ at low $q$, and (iii) a relatively low difference between positive and negative charges that tends to promote a lower $S(0)$, typical of repulsive-dominated interactions. It is also important to realize that these candidates were obtained from a dataset that did not include the exact features we then obtained as output (see SI), underlying even more the soundness of our approach and thus the possibility of employing an initial dataset that only roughly resemble the expected antibody features. 

In order to verify whether this result can be further refined, we enlarge the training dataset by considering ten additional pairs of structure factors, for each of the two concentrations. We choose to create this additional piece of training dataset by randomly varying the sum of positive and negative charges as compared to the best results by $\pm 2$ (see also Methods). Since the goal is to obtain the best agreement with experimental data, this should direct the outcome even more towards a result that is closer to real, while still having the generality introduced with the first optimization cycle. The strategy proves successful since, as shown in Figure~\ref{fig:optimizationml}(a.II), the additional training dataset fully includes the target, and the three best outcomes of the second optimization cycle generate configurations with an overall MSE that is lower than in the first cycle (see Figure~\ref{fig:optimizationml}(b.II)). The third additional training dataset is then built in a similar way, by generating additional configurations and running corresponding simulations for variations of the sum of positive and negative charges by $\pm 1$ with respect to the best outcome of the second optimization cycle, again yielding even lower values of the MSE.
We emphasize, however, that the first cycle alone is sufficient to achieve qualitative agreement, thereby avoiding the need for large number of simulations when fast scanning is desired.

\begin{figure*}[t!]
\centering
\includegraphics[width=\textwidth]{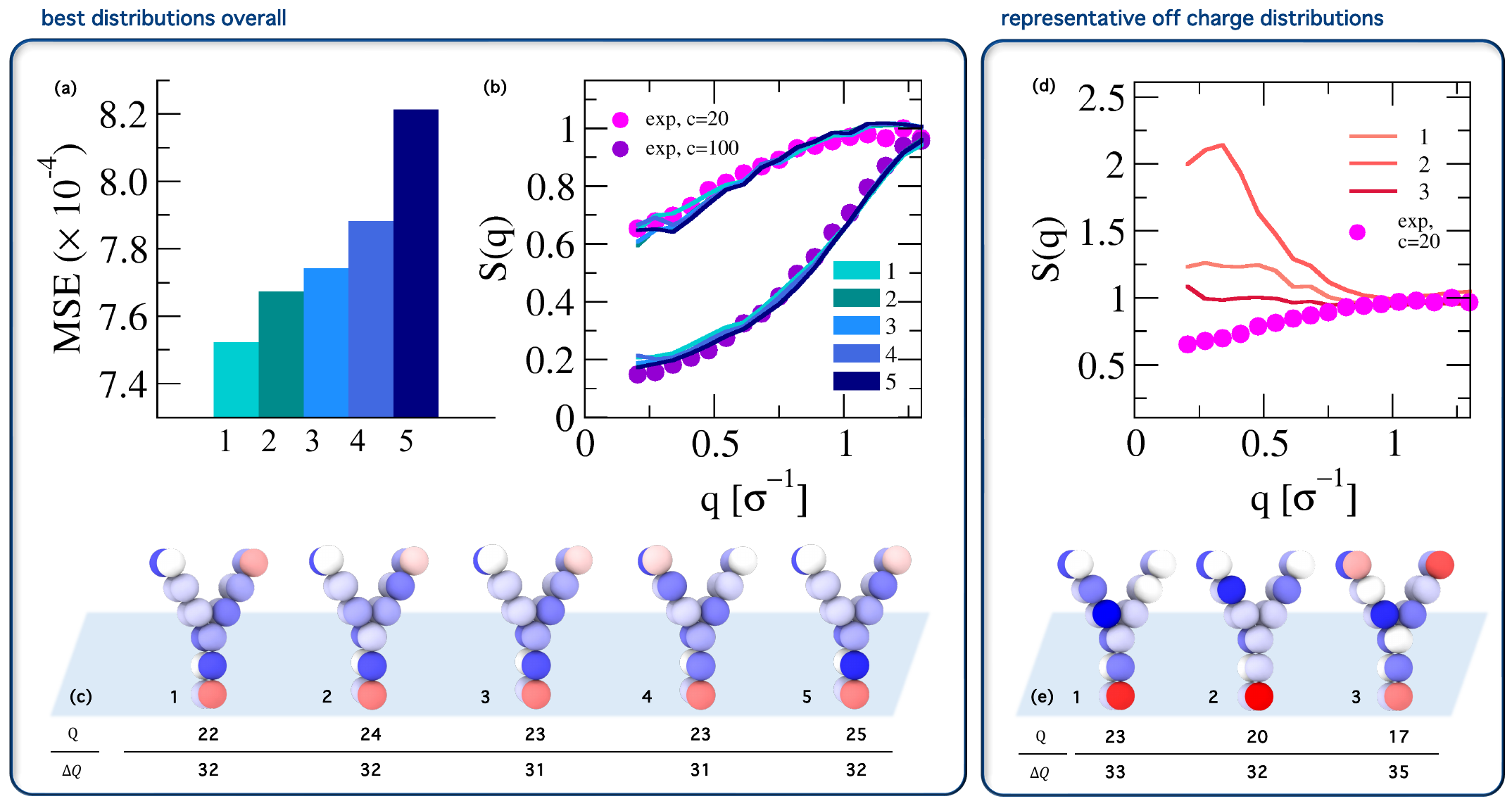}
\caption{\small \textbf{Best (and off) charge distributions.} (a) Mean square error (MSE) and (b) static structure factors $S(q)$ as a function of the scattering vector $q$ for antibody concentrations $c=20$ and $100$ mg/ml, and (c) corresponding snapshots for the best five charge distributions, both from specifically-created configurations used for the training datasets and from output configurations of the NN. (d) Representative static structure factors $S(q)$ as a function of the scattering vector $q$ for an antibody concentration $c=20$ mg/ml, and (e) corresponding snapshots for representative charge configurations whose charge features have a non-satisfactory agreement with the experimental $S(q)$. Symbols in (b) and (d) are for experimental data, taken from Ref.~\cite{camerin2026beyond}.}
\label{fig:bestoverall}
\end{figure*}
 
In this way, we are able to create a final ranking of the best antibody representation, including both the configurations used as training dataset and the outcomes of the different optimizations cycles. This is reported in Figure~\ref{fig:bestoverall}(a,b) where we show the MSE for the five best charge distributions and the corresponding static structure factors for both concentrations. Additional data for two other concentrations in comparison with experiments are shown in the SI.
Ultimately, it appears that the best configurations that match experimental outcomes for both concentrations analyzed have the following features: $22 \le Q \le 25$, $31 \le \Delta Q \le 32$, $\mathcal{Q}={\rm ``tips"}$. The agreement is in all cases remarkable with MSEs $<10^{-3}$ and structure factors showing only minimal deviations compared to the target. By looking at the charge distributions reported in Figure~\ref{fig:bestoverall}(c), we observe that the presence of negative tips is indeed preserved in all cases. Furthermore, it appears that having either a negatively charged or a neutral bead on the upper extremities, as well as varying the distribution of the positive charges in the central part of the antibody, only marginally affects the solution structure. 
Similarly, small variations of the overall negative charge ($-4$ or $-5$) on side $A$ do not notably change the $S(q)$.

An important aspect to underline relates to the way the overall negative charge is distributed on the antibody. In Figure~\ref{fig:bestoverall}(d), we report the $S(q)$ for cases with similar charge features to the ones that match the experimental data, but where the negative charge is concentrated in one of the three tip beads; we also show the effect of a more marked variation of the best-matched parameters. The corresponding snapshots and relevant features are shown in panel (e). It is evident that negative charges concentrated in a single point on an antibody with a net positive overall charge create a strongly attractive patch that originates strongly correlated structures, which can be detected as a substantial increase in $S(0)$. Our protocol is therefore autonomously able to distinguish between these cases and to establish that the best representation of the charge distribution on the antibody under investigation has negative attractive sites evenly distributed on the tips of the Y-shaped objects. We also note that these kinds of charge distributions would not be easily guessed from the electrostatic potential surfaces depicted in Figure~\ref{fig:provisionalmodel}(b).
\\
\\
\indent Thus far, our approach has relied on minimal prior knowledge of the antibody’s characteristics to generate an initial, plausible set, aimed at converging toward a coarse-grained model that faithfully represents the structure factors of the experimental antibody under investigation. While this strategy provides a solid foundation, a deeper understanding of which specific parameters most significantly influence $S(q)$ is essential for exerting meaningful control over the system. To this end, we introduce the SHAP (SHapley Additive exPlanations) algorithm~\cite{lundberg2017unified,lundberg2020local}, which enables us to quantify the individual contributions of the various electrostatic parameters to variations in $S(q)$. In essence, SHAP identifies the features that drive the largest changes in the solution structure, highlighting the parameters that should be prioritized when tuning solution properties. Some of these features may overlap with those initially used for sampling the parameter space, others that we specifically add, either global or more local properties,  may emerge as relevant for understanding the system's behavior. 

\begin{figure}[t!]
\centering
\includegraphics[width=0.5\textwidth]{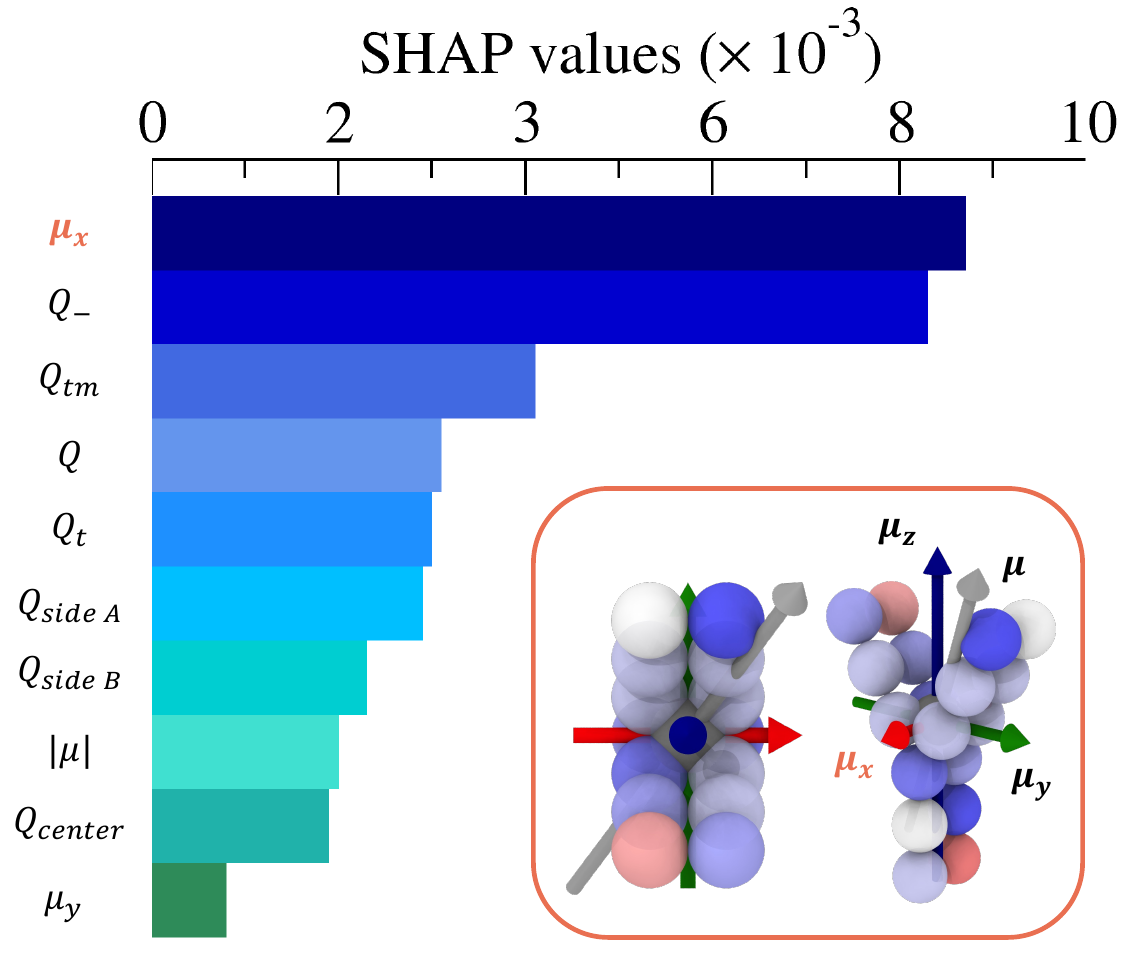}
\caption{\small \textbf{Relevant physical features.} Ten best physical features of the antibodies that determine the biggest variations in the solution structure according to the SHAP algorithm. In order of relevance: $x$-component of the dipole moment $\mu_{x}$, sum of the negative charges $Q_-$, sum of the middle and tips charges $Q_{tm}$, total charge $Q$, sum of the charges located at the tips $Q_t$, sum of the charges located at side $A$ of the antibody $Q_{\rm side \ A}$, sum of the charges located at side $B$ of the antibody $Q_{\rm side \ B}$, modulus of the dipole moment $|\mu|$, sum of the charges of the central beads at side $A$ of the antibody $Q_{\rm center}$, $y$-component of the dipole moment $\mu_y$. The inset shows the antibody whose charge distribution produces the lowest mean squared error among the best overall configurations together with the vectors of the center-of-mass total dipole moment and the respective $x,y,z$ components.}
\label{fig:shap}
\end{figure}

Therefore, based on our physical intuition, we create a set of features that might be relevant for describing the charge state of the antibody. Among others, we include the sum of positive and negative charges, the dipole moment calculated from each charge distribution at its center of mass, and the product between charges in different beads of the antibody. The complete list of features is provided in the SI. Then, after having trained a machine learning model that relates all the features to the corresponding structure factor, we apply SHAP to the entire dataset we have generated and obtain, for each of the features, the corresponding ``SHAP value" which basically quantifies the contribution of each input feature. More details on the overall implementation of the algorithm are reported in Methods. The extracted SHAP values are reported in Figure~\ref{fig:shap} for the ten most important features identified. The outcome of this analysis is remarkable and reveals that the $x$-component of the dipole moment of the antibody $\mu_x$ is the most prominent feature in determining a variation of the solution structure. To provide a physical understanding of this feature, we display the center of mass dipole moment of the antibody and the respective components for the charge distribution associated with the lowest MSE as an inset of Figure~\ref{fig:shap}. We thus observe that the $x$-component is the one associated to the charge distribution \textit{across} the two layers of which the antibody model is made. By looking at the $x$-component of the dipole moment for all the antibodies selected as plausible in Figure~\ref{fig:bestoverall}, we indeed observe that they always have $2.5 \lesssim \mu_x/(e^*\sigma) \lesssim 3.0$. 
In other words, it appears that to have control on the static solution properties of the antibody under investigation the choice of the $x$-component of the dipole moment is crucial. 
We further note that the sum of the negative charges $Q_-$ and the sum of the charges of tips and middle beads $Q_{tm}$, also play a relevant role in determining $S(q)$. These, in fact, remain consistent features among our five best representations of the antibody.

\subsection*{Prediction of additional solution properties}

At last, after having established that a narrow set of charge patterns reproduces the experimental $S(q)$ for different concentrations, and having identified the key electrostatic descriptors controlling these trends, we now test whether this optimized 18-bead coarse-grained model is also capable of reproducing other solution properties that are commonly measured experimentally. In particular, to establish a direct connection between the microscopic interaction landscape and the experimentally observed solution properties, we combine simulation and experimental results with analytical calculations from liquid-state theory. In Methods, we describe the main concepts used here.

\begin{figure}[t!]
\centering
\includegraphics[width=0.5\textwidth]{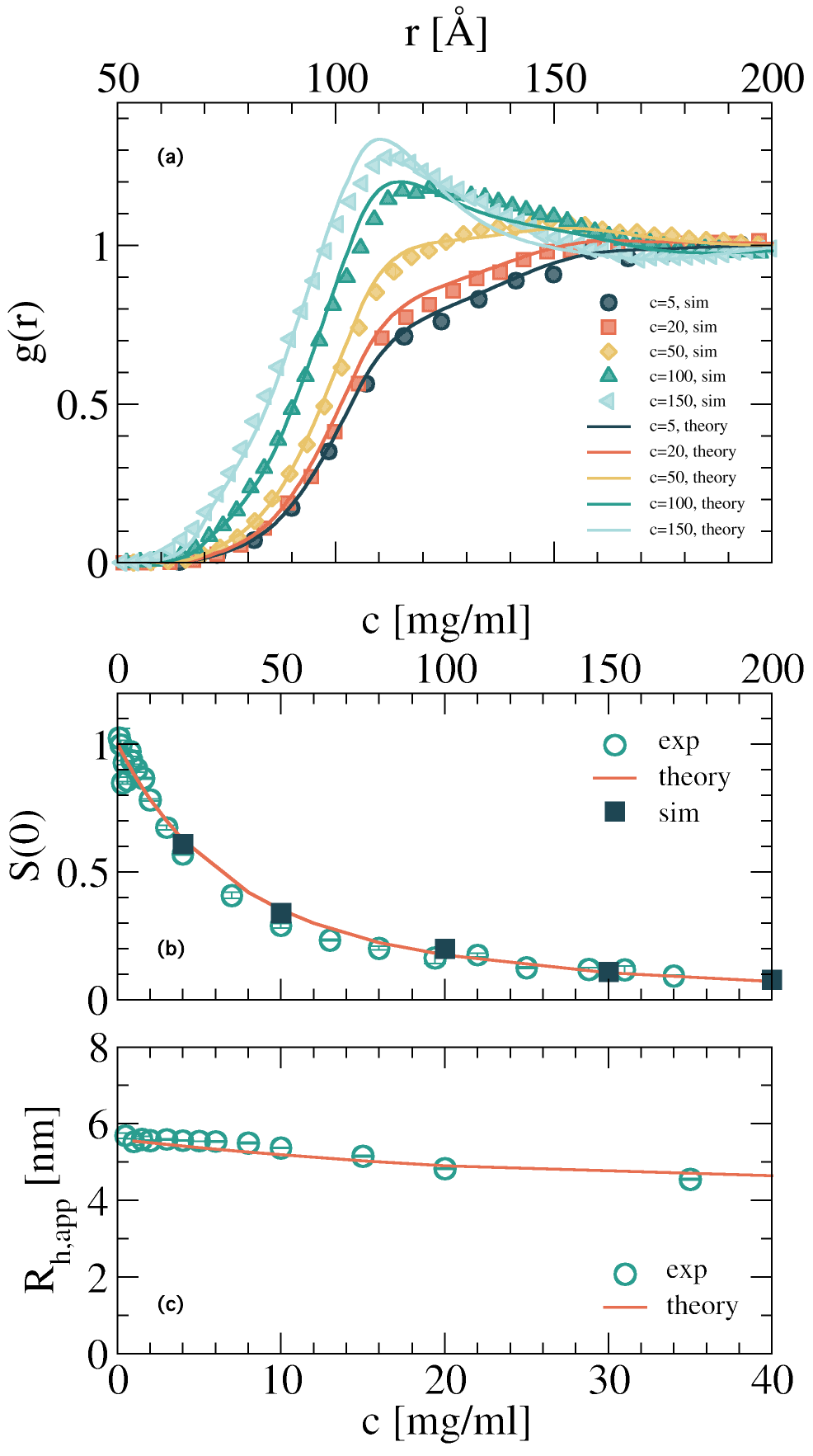}
\caption{\small \textbf{Comparison to liquid-state theory predictions.} (a) Center-of-mass radial distribution function $g(r)$ as a function of the distance $r$ as obtained from ensemble simulations of the 18-bead model (symbols) and calculated from the theory (lines) for different antibody concentrations $c$. (b) Low-$q$ limit of the static structure factor $S(0)$ as a function of antibody concentration $c$ as obtained from the theory (line), static light scattering experiments (circles) and simulations (squares). (c) Apparent hydrodynamic radius $R_{h,app}$ as a function of antibody concentration $c$ as obtained from the theory (line) and dynamic light scattering experiments (circles). Light scattering data in (b) and (c) are taken from Ref.~\cite{camerin2026beyond}.
}
\label{fig:theory}
\end{figure}

We start from the ensemble simulations with explicit ions, from which we extract the radial distribution functions  of the antibodies' center of mass $g(r)$ up to low concentrations such as $c=5$ mg/ml. We focus on the charge distribution that give rise to the highest agreement with experimental data, labeled as $1$ in Figure~\ref{fig:bestoverall}. These radial distribution functions encode the microscopic organization of the system and reflect the balance between short-range attractive and excluded-volume effects and long-range electrostatic repulsive and attractive interactions. Exploiting the relation between structure and protein-protein interactions, it is possible to extract an orientation-averaged potential of mean force $U(r)$ at low concentrations through a Boltzmann inversion (BI) method, which relies on the ``inversion" of the radial distribution function $g(r) \approx$ exp$[-\beta U(r)]$. We subsequently use $U(r)$ as an input to solve the Ornstein–Zernike equation within the hybrid mean-spherical approximation (HMSA). The comparison between the radial distribution function extracted from simulations and the ones calculated using HMSA is reported in Figure~\ref{fig:theory}(a). It shows that the theoretical treatment based on a potential of mean force obtained from an ensemble simulation with the optimized 18-bead structure at low concentration is able to capture the correct antibody distribution in solutions for the entire range of concentrations analyzed. Importantly, not only the position of the peaks of the $g(r)$ is correctly captured, but we also confirm the quite unusual shape of $g(r)$ at intermediate distances for low concentrations as an intrinsic feature of the system.

We next translate the same interaction picture into the quantities that are most directly compared against experiments. From the HMSA solution we compute the structure factor and, in particular, its low-$q$ limit $S(0)$, which is directly related to the osmotic compressibility. Figure~\ref{fig:theory}(b) reports $S(0)$ as a function of concentration, comparing the theoretical prediction to both simulation estimates and experimental data discussed previously. The agreement is remarkable at all concentrations, especially in the dilute regime, where $S(0)$ is most sensitive to the net balance between repulsion and any residual attraction encoded in the effective interaction. This shows that the potential of mean force extracted from the explicit-ion simulations carries the correct “integrated” strength and range of interactions needed to reproduce the experimentally observed compressibility without further adjustments. 

Finally, we extend the comparison to dynamical observables, using the fact that the same pair correlation functions that govern the thermodynamics also set the strength of hydrodynamic coupling in solution. In this way, following the procedure described in Methods, we combine the large-scale hydrodynamic response $H(0)$ with $S(0)$, to obtain the concentration dependence of the hydrodynamic radius $R_{h,app}$ measured with dynamic light scattering (DLS). This is shown in Figure~\ref{fig:theory}(c) together with experimental values. Also in this case, the model reproduces the measured concentration dependence in the low-concentration regime, indicating that the effective interaction extracted from the explicit-ion simulations is not only consistent with the static correlations but also captures the magnitude of the acceleration of collective dynamics observed experimentally.

This “end-to-end” agreement is important because it shows that the inferred interaction picture is not merely a convenient parametrization of the solution structure, but captures the underlying pair correlations that control both thermodynamics, via $S(0)$, and collective dynamics, via $H(0)$ and $R_{h,app}$. As a result, the 18-bead framework provides a compact route to connect molecular charge anisotropy to experimentally accessible macroscopic properties, while retaining an explicit representation of ions that is essential when charge patchiness and ion–protein correlations play a dominant role~\cite{camerin2026beyond}. Importantly, this level of coarse-graining does not come at the expense of accuracy. On the contrary, by comparing the same quantities obtained from a more resolved, amino acid model in a recent work by some of us~\cite{camerin2026beyond}, it actually constitutes a substantial improvement, due to the computational difficulties in obtaining a reliable potential of mean force for such complex simulations. This is now out-performed by the current coarse-grained models where charge anisotropy has been successfully deciphered.

\section*{Conclusions}

Our results demonstrate that correctly describing the heterogeneous and anisotropic charge distribution of antibodies is crucial to understand and reproduce their collective behavior in solution. Even subtle variations in charge parameters can produce marked changes in the overall structure of the solution. By first inspecting the amino acid  charge distribution and the associated electrostatic isopotential surfaces, we obtained a clear physical intuition of how charges are organized and how they might drive mesoscale interactions. This guided the construction of a 18-bead “baseline” coarse-grained model, for which an initial guess of the coarse-grained charged distribution could be drawn. In this way, the emergent properties captured by the model remain anchored to the realistic structural characteristics of the protein under investigation.
More generally, this approach underlines a key design principle: analyzing the molecular charge landscape \textit{a priori} can provide valuable guidance for constructing coarse-grained models of biomolecules or heterogeneously charged colloids. 
Building on this physically-informed baseline, we introduced a machine-learning-assisted inverse-design strategy that uses experimental structure factors at multiple concentrations as direct constraints to reconstruct effective charge distributions, circumventing the limitations of a trial-and-error parametrization. 
Importantly, the optimization converges efficiently even with relatively small training datasets, demonstrating that physically-guided coarse-graining and data-driven refinement are complementary rather than competing strategies. We emphasize that this strategy is broadly applicable, ranging from highly anisotropic to nearly spherical molecules, where shape may be simpler but the charge landscape can still be markedly heterogeneous~\cite{gnidovec2025anisotropic}.

Crucially, the resulting coarse-grained description is not limited to reproducing scattering signatures. When combined with explicit-ion simulations and liquid-state theory, it yields a coherent and quantitatively consistent description of independent observables, including osmotic compressibility and hydrodynamic response. This internal consistency shows that the inferred interaction landscape captures the essential pair correlations governing both thermodynamics and collective dynamics at different concentrations, rather than merely fitting a single experimental quantity.
At the same time, the coarse-grained charge distributions we identify should be interpreted as \textit{effective representations}. In fact, antibodies are not rigid objects~\cite{castellanos2016role,girelli2021molecular,biehl2025diffusion}; their three domains are connected by a hinge region that allow substantial flexibility, implying that charge exposure is inherently dynamic, with attractive patches being transiently screened or enhanced depending on molecular orientation and solution conditions. While the present framework already captures the dominant electrostatic features controlling the solution structure, explicitly incorporating conformational variability into future coarse-grained descriptions -- through semiflexible linkers, mixed conformational ensembles, or hybrid multiscale strategies -- could provide additional insights into how charge heterogeneity and anisotropy couples to structural dynamics in concentrated solutions.

The colloidal framework adopted here proved to be particularly powerful: antibodies were treated as anisotropic charged colloids whose mesoscale properties can be directly traced back to their spatial charge arrangement. 
This viewpoint also underscores the parallels between protein solutions and classical charged colloids, where patchy interactions dictate thermodynamics and phase behavior~\cite{stradner2020potential,bianchi2017limiting}. Besides, our approach aligns with the growing innovative field of inverse design in soft matter physics, where collective properties are engineered by tuning particle-level interactions. Recent advances in colloidal science have demonstrated the potential of this paradigm~\cite{coli2022inverse,wang2025inverse,beneduce2025inverse,liu2024inverse,rivera2023inverse,ma2019inverse,russo2022sat,gnidovec2025controlling,dijkstra2021predictive,curtis2025predicting}.
Here, we establish inverse design of charge patterns as a predictive framework to connect experimental scattering data to underlying amino acid-scale features. Complementing this view, our SHAP-based feature attribution analysis reveals that the dipole moment emerges as an additional molecular-level descriptor with significant influence on the solution structure. 
In this context, it is worth noting that machine-learning approaches applied to antibodies have so far been predominantly used to address single-molecule problems, such as antigen–antibody recognition or sequence-level optimization~\cite{tiller2015advances,chan2025fifty}. Here, instead, machine-learning is used to infer how molecular charge distributions manifest at the level of collective behavior in solution, by linking experimental scattering data directly to effective intermolecular interactions and ensemble-averaged properties. This perspective also distinguishes our work from earlier studies that related antibody electrostatics to experimental measurements~\cite{ferreira2019electrostatically,singh2014dipole,roberts2014role,Yadav2012}, by uncovering the direct link between molecular charge features and solution properties.

Finally, our findings have direct implications for therapeutic formulation and developability assessment. 
Most antibodies are required in concentrated formulations for subcutaneous administration, where intermolecular interactions strongly affect stability, compressibility, and transport properties~\cite{filipe2012immunogenicity,lundahl2021aggregation,pham2020protein,mosca2025multiscale}. In this context, our analysis provides a concrete, physically-grounded interpretation of how localized charge patterns, and particularly the presence and placement of negative patches, shape effective interactions and thereby control measurable solution behavior. For the present case, the results highlight that negatively charged regions at the tips act as dominant interaction motifs, and that their spatial distribution, rather than net charge alone, can decisively alter the collective response. At the same time, these regions may be coupled to functional constraints such as antigen recognition, raising the broader design challenge of preserving activity while avoiding charge localization that promotes undesirable intermolecular association. By framing the problem in terms of effective interactions constrained by scattering and validated by thermodynamic and hydrodynamic observables, our approach provides a rational basis for navigating this trade-off. 
More broadly, this framework lays the basis for the next generation of formulation-relevant modeling strategies. In future developments, validated interaction landscapes could serve as a starting point for increasingly refined coarse-grained descriptions capable of addressing collective slowdowns and rheological trends at high concentrations. Such extensions may ultimately contribute to a more predictive understanding of viscosity increases, which remains one of the main practical bottlenecks for injectable antibody formulations~\cite{mijangos2025developing}.

In summary, this work provides a generalizable pathway linking molecular charge anisotropy to collective solution properties.
By combining colloidal coarse-graining, explicit electrostatics, liquid-state theory, and machine learning optimization, we establish a versatile framework that advances fundamental understanding of electrostatics in complex biomolecules and offers a predictive tool for controlling the behavior of heterogeneously charged soft matter.

\footnotesize
\section*{Methods}

\noindent \textbf{Materials.} We use the monoclonal antibody Cetuximab, or Erbitux, a recombinant chimeric human/mouse IgG1 monoclonal antibody that binds to the epidermal growth factor receptor, as a model system for an antibody with heterogeneous charge distribution. A detailed experimental characterization of its solution properties has recently been reported~\cite{camerin2026beyond}. The amino acid sequence for both the light and heavy chains of Cetuximab are reported in the IMGT database (www.imgt.org) and the drug bank (www.drugbank.ca), and its primary sequence has also been reported by Dubois et al.~\cite{Dubois2008}. The crystal structure of the antigen binding fragment (Fab) has been reported by Li et al.~\cite{Li2005} and is referenced in the RCSB Protein Data Bank. The atomistic structure of the antibody was built by homology modeling using the Molecular Operating Environment (MOE) software (Chemical Computing Group, Inc.).

\noindent \textbf{Experimental characterization.} We use previously reported experimental results for this antibody~\cite{camerin2026beyond}. There,  experimental measurements were made using antibody solutions in a low ionic strength buffer (10 mM L-histidine at pH = 6.0) at an ionic strength of 7 mM. Static and dynamic light scattering were used to determine the osmotic compressibility or the low-$q$ limit of the static structure factor $S(0)$ and the apparent hydrodynamic radius $R_{h,app}$ as a function of the concentration $c$. Small-angle X-ray scattering was used to determine the static structure factor $S(q)$. For details of the sample preparation and the experiments, see Ref.~\cite{camerin2026beyond}.

\noindent \textbf{Electrostatic isopotential surface.} The charge distribution of the antibody under investigation is better visualized by means of the so-called electrostatic isopotential surfaces. 
From the amino acid representation of the antibody obtained in Ref.~\cite{camerin2026beyond} these are calculated by using the Adaptive Poisson-Boltzmann Solver (\textsc{apbs})~\cite{jurrus2018improvements} embedded in the Visual Molecular Dynamics (\textsc{vmd}) software~\cite{humphrey1996vmd}.

\noindent \textbf{Antibody coarse-grained model.} For modeling the antibody, we build on the model introduced by some of us in Ref.~\cite{camerin2025electrostatics}, where a 9-bead model inspired by anisotropic colloids was employed for a different antibody. For the present antibody, we increase the number of beads due to its complex charge distribution. We thus resort to a 18-bead model with two structurally identical 9-bead layers, as the previous one, placed in contact on top of each other (see Fig.~\ref{fig:provisionalmodel}(c)).
Each bead has the same diameter $\sigma$, the unit of length in simulations, and a unit mass $m$. The layers thus consist of 9-beads each (labeled as $A$ and $B$ sides, see Fig.~\ref{fig:provisionalmodel}), each arranged in a Y-shaped molecule whose beads form angles of $150^{\circ}$ and $60^{\circ}$ with the others. Importantly, the geometry of each antibody is fixed and they are treated as rigid bodies.

\noindent \textbf{Interaction potentials for the coarse-grained model.} Each bead of the antibody coarse-grained model is assigned a charge $\tilde{q}_{i}$; $i,j$ beads interact with each other via the Coulomb potential
\begin{equation}
V_{Coul} (r_{ij})= \frac{\tilde{q}_i \tilde{q}_j\sigma}{e^{*2}r_{ij}}\epsilon
\end{equation}
where is $r_{ij}$ is the distance between the beads, $e^{*}=(4 \pi \epsilon_0 \epsilon_r \sigma \epsilon)^{1/2}$ is the reduced charge unit with $\epsilon_0$ and $\epsilon_r$ the vacuum and relative dielectric constants, and $\epsilon$ the energy unit. For an efficient treatment of the long-range nature of the electrostatic interactions we employ the particle-particle particle-mesh (PPPM) solver~\cite{hockney2021computer}. 
We then account for excluded volume and van der Waals interactions by using a modified Lennard-Jones (mLJ) potential 
\begin{equation}
V_{\rm mLJ}(r_ij)=
\begin{cases}
C\gamma\left[\left(\frac{\sigma}{r_{ij}}\right)^{96}-\left(\frac{\sigma}{r_{ij}}\right)^{6}\right] & \text{if $r_{ij} \le 3.5\sigma$}\\
0 & \text{otherwise,}
\end{cases}
\end{equation}
with $\gamma=1.28\epsilon$ and $C=0.1$ an adjustable prefactor.

\noindent \textbf{Simulation details and parameters.} We run equilibrium molecular dynamics simulations in the NVT ensemble, fixing the number of antibodies $N=100$. Simulations are performed in a cubic box with periodic boundary conditions in the three dimensions. The volume is adjusted to reproduce different values of the experimental antibody concentration $20 \leq c/(\rm mg \, ml^{-1})\leq 200$, as detailed in Ref.~\cite{polimeni2024multi}. For all concentrations analyzed, we use $\sigma=2.55$ nm. 

The system is ensured to be charge neutral by adding a congruent number of positive and negative monovalent counterions ($\tilde{q}=\pm 1$), which also interact via $V_{Coul}$. To account for their excluded volume, we use a simple Weeks-Chandler-Anderson (WCA) potential,
\begin{equation}
V_{\rm WCA}(r)=
\begin{cases}
4\epsilon\left[\left(\frac{\sigma_{ion}}{r}\right)^{12}-\left(\frac{\sigma_{ion}}{r}\right)^{6}\right] + \epsilon & \text{if $r \le 2^{\frac{1}{6}}\sigma_{ion}$}\\
0 & \text{otherwise,}
\end{cases}
\end{equation}
with $\sigma_{ion}=0.1 \sigma$, as in a previous work by some of us~\cite{camerin2025electrostatics}.
In the simulations, we also consider the ions coming from the buffer with a concentration $c_{m,buffer}=7$ mM. To this aim, we determine the number of particles to be added in the simulation box by calculating the molar concentration of the antibodies $c_{m,mAb}$ in solution, considering the antibody molecular weight $M_w=152000$ g/mol. The number of buffer ions to be added is thus determined as $N_{buffer}=2N \frac{c_{m,buffer}}{c_{m,mAb}}$, half of which is assigned a $+1$ charge and the other half a $-1$ charge.

We set the reduced temperature $T^*=k_BT/\epsilon=1.0$, with $k_B$ the Boltzmann constant and $T$ the temperature by means of a Langevin thermostat. Equilibration runs are carried out for at least $1\times10^6 \delta t$, with $\delta t=0.002\tau$ and $\tau=\sqrt{m \sigma^2/\epsilon}$ the unit of time. A subsequent production run is carried out for at least $2\times10^7 \delta t$. Coarse-grained molecular dynamics simulations are run with \textsc{lammps}~\cite{thompson2022lammps}.

\noindent \textbf{Measured quantities.} We calculate the static structure factors  as
\begin{equation}
S(q)=\frac{1}{P(q)}\frac{1}{18N}\left\langle\left(\sum_i^{18N}\sin \boldsymbol{qr}_i\right)^2+\left(\sum_i^{18N}\cos \boldsymbol{qr}_i\right)^2\right\rangle
\label{eq:sq}
\end{equation}
with the sums running over all $18N$ beads of the system, $P(q)$ the form factor of a single 18-bead antibody model, $\boldsymbol{q}$ the scattering vectors, $\boldsymbol{r}_{i}$ is the positions of the $i$-th bead, and $\langle \cdots \rangle$ indicating an average over different configurations of the system and over different orientations of the scattering vectors $\boldsymbol{q}$.

The dipole moment of each antibody is calculated in magnitude as
\begin{equation}
\mu = \sum_{i=1}^{18} \tilde{q}_i r_i.
\end{equation}
As detailed in the text, the dipole vector is consistently calculated using a fixed orientation of the antibody: the $zy$ plane lies between the two layers of the antibody, the $xz$ plane separates the two upper arms of the Y-shaped structure, and the $z$-axis is aligned with the direction defined by the lower arm.

For quantifying the agreement between the static structure factors obtained from simulations and the experimental ones, during the optimization cycles, we make use of the mean squared error (MSE). This is defined as
\begin{equation}
    MSE=\frac{1}{N_q}|\mathbf{S}-\mathbf{S}_{\mathrm{exp}}|^2,
\end{equation}
where $\mathbf{S}_\mathrm{exp}$ is defined as the vector obtained by concatenating the experimental $S(q)$ at $c=20$ and $c=100$ mg/ml, while $\mathbf{S}$ is the corresponding vector from simulations and $N_q$ denotes the total number of $q$ scattering vectors considered.

\noindent \textbf{Preparation of the training datasets.} The training datasets consist of a series of static structure factors $S(q)$ obtained by running molecular dynamics simulations of ensembles of antibodies each with a different combination of charges. 
In particular, we choose to vary the overall total charge $Q$, the difference between the sum of positive and negative charges $\Delta Q$, and the positioning of the negative charges on the structure of the antibody $\mathcal{Q}$ giving rise to tips, middle, tips and middle, and spread configurations, as described in the main text. The choice of the explored range of $Q$ and $\Delta Q$ is based on a rough overlap between the amino acid representation of the antibody and the coarse-grained model, which allowed us to perform a mapping of the charges from the finer to the more coarse-grained representation, as previously done in Ref.~\cite{camerin2025electrostatics}. Based on this first initial guess and knowing that the total charge extracted from amino-acid simulations was found to be $\approx +24$, we pick $Q=20,24,28$ and $\Delta Q=30,35,40$. We recall that the values of charges are given in units of $e^*$, as discussed above. 
The overlap between configurations also provided a first initial guess for side $B$ of the antibody, whose beads are all positively charged or neutral.

We automate the assignment of the charges on each bead by means of a simple algorithm. For a given set of parameters, negative charges are assigned based on the specific charge distribution $\mathcal{Q}$. All remaining beads are required to be non-negative. 
We note that this affects only side $A$. The total amount of negative charges is then determined by satisfying \textit{at best} the predetermined conditions of $Q$ and $\Delta Q$. In case both conditions cannot be met simultaneously, one of the two is relaxed. We also allow for a variation in the positive or neutral charges initially assigned by $\pm 2$ units of charge. Since in the initial overlap between differently coarse-grained configurations none of the charges is found to have a higher absolute value than $6$, we also impose that none of the bead can retain a charge higher than this value. Subsequently, for including additional variability in the dataset, we randomize the charges on side $A$ by still requiring that the sum of positive and negative charges is kept fixed, and by leaving the positioning of the negative charges unchanged as compared to the initially set distribution $\mathcal{Q}$. The complete initially-created dataset is reported in the SI.
Importantly, this strategy allows for the systematic and efficient generation of charge distributions that comprehensively cover the selected parameter space. Although this training set remains inherently limited and does not resolve the charge parameters at a very fine scale, its ability to produce satisfactory outcomes even in the first round of optimization (see main text) indicates that similarly sized sets, differing slightly in parameter assignments, are likely to yield comparable performance.

The charge distributions used for enlarging the training dataset at the second and third optimization cycles are instead obtained on the basis of the main physical features extracted as outcome of the first and second optimization cycle, respectively. While in both cases we use charge distributions with negative charges at the tips of the antibodies ($\mathcal{Q}_t$), in the former we allow for a variation of the sum of positive and negative charges by $\pm 2$ and in the latter by $\pm 1$ as compared to the best outcome in the previous respective cycles. The resulting total charge is additionally spread randomly over the beads, always without varying the position of the negative charges and not allowing beads with absolute value of the charge higher than $6$.

\noindent \textbf{Neural Network architecture.} To unveil the charge distribution that generates the experimental structure factor at different concentrations, we employ supervised machine learning techniques. In particular, we build a (deep) neural network (DNN, or just NN) that learns the charge distribution from the low-$q$ region of the structure factor $S(q)$, obtained from molecular dynamics simulations. 

The input dataset consists of a set of discretized $S(q)$ 
corresponding to various charge distributions at the two concentrations, $c_1=20$ and $c_2=100$ mg/ml. First, we spline the $S(q)$ for both concentrations and build the input vector $\mathbf{S}$ for each distribution $i$. Each entry $\mathbf{S}_{i,j}$ is constructed by evaluating the splines at regular intervals of $\Delta q=0.07 \sigma^{-1}$ starting from $q\sigma=0.2$ up to $q\sigma=1.3$. The first $17$ entries correspond to the $S(q)$ at $c=20$ mg/ml, followed by the values for $c=100$ mg/ml. The output vector $\mathbf{Q}$ contains the respective charge distribution of the two-layer antibody, where each entry $j$ of $\mathbf{Q}_{i,j}$ corresponds to the charge at a fixed bead position. 

Given the limited amount of training data, and with the aim of constructing a framework that is capable of predicting the charge distribution from the experimental data, we use a NN architecture with three hidden layers, each employing the ReLU activation function~\cite{sharma2017activation}. The first two have $64$ neurons each, followed by the third layer with $36$ neurons. The input layer receives a $34$-dimensional vector, while the output gives a vector with dimension $18$ representing the charge of each monomer. To reduce the risk of overfitting and to ensure faster convergence, we incorporate batch normalization~\cite{ioffe2015batch} and dropout in each layer. The first standardizes the input of each layer to have zero mean and a standard deviation of $1$. The latter randomly sets a fraction $p$ of the neurons to zero during the training. In our case we fix $p=0.2$. 

The NN is trained by minimizing a loss function $\mathcal{L}$ using Adam optimization~\cite{kingma2014adam}, with a learning rate of $0.001$ and weight decay of $1 \times 10^{-4}$. From the amino acid simulations, the total charge $Q$ is known to roughly lie between $20$ and $28$. Therefore, we build a loss function $\mathcal{L}$ that penalizes configurations with a total charge outside this range, 
\begin{equation}
  \mathcal{L}  =\frac{1}{M}\sum_{i=1}^{M} \left| \mathbf{Q}_i-\mathbf{Q}_i^{out}\right|^2+\lambda \left(\mathrm{max}(0,20-\mathbf{Q}_{i}^{out})^2+\mathrm{max}(0,\mathbf{Q}_{i}^{out}-28)^2 \right),
    \label{eq:error}
\end{equation}
where $M$ is the total number of samples in the training data set,  $\lambda=5$ is the total charge penalty, and $\mathbf{Q}_{i}^{out} = \sum_j \mathbf{Q}_{ij}^{out}$ is the predicted total charge for sample $i$. The NN is trained for at least $2 \times 10^4$ epochs.

\noindent \textbf{SHAP algorithm.} To understand the most relevant features that contribute to the low-$q$ behavior of $S(q)$, we employ an explainable machine learning approach named SHAP (SHapley Additive exPlanations)~\cite{lundberg2017unified,lundberg2020local}, which uses Shapley values from cooperative game theory~\cite{winter2002shapley}. The main goal of this approach is to quantify the average marginal contribution of each feature to a prediction from a machine learning model. 

To do so, we construct a dataset using the $S(q)$ and the features that characterize the charge distribution of the data from molecular simulations. Specifically, we use an input vector $\mathbf{F}_i$ that includes the total charge $Q_{\mathrm{tot}}$, the sum of all positive charges $Q_{+}$, the sum of negative charges $Q_{-}$, their difference $\Delta Q=Q_+-Q_-$, as well as the sum of the charges located at the tips $Q_{\mathrm{t}}$, middle $Q_{\mathrm{m}}$, center $Q_{\mathrm{c}}$, the three components of the dipole moment ($\mu_x,\mu_y,\mu_z$), its magnitude $\mu$, and a set of 9 charge-products that encode neighboring charge combinations for distribution $i$. The output vector $\mathbf{S}_i$ corresponds to the discretized $S(q)$ at a specific concentration. 

We then train a separate Random Forest Regressor (RFR)~\cite{ho1995random} for each $q$-value of the $S(q)$. An RFR is a machine learning model that combines the predictions of a number of decision trees, each trained on a random subset of the data. Each tree recursively splits the data into smaller subsets based on conditional rules until a maximum tree depth is reached. The final prediction is obtained by averaging the outputs of all the trees, which improves accuracy and reduces overfitting. Here, we use $100$ trees per forest, each with a maximum depth of $10$. After training the RFR, we apply the SHAP algorithm~\cite{lundberg2020local} to compute the Shapley values per $q$ in the $S(q)$. We then average the values  across all $q$-points and rank the features by their overall contribution, identifying those that most strongly influence the shape of the structure factor.

\noindent \textbf{Calculation of pair distribution functions, osmotic compressibility and apparent hydrodynamic radii.} We use liquid state theory to calculate the pair distribution function $g(r)$ and the osmotic compressibility or $S(0)$ as a function of concentration for the given potential of mean force obtained from the low concentration ensemble simulation~\cite{Goodstein1975,Naegele1996,cipelletti2025interacting}. The starting point is the link between the orientationally-averaged static structure factor and the pair distribution function $g(r)$ given by,
\begin{equation}
	\label{pair-distribution}
	S(q) = 1 + 4 \pi \rho   \int_{0}^{\infty} r^2\left[g(r) - 1\right] \frac{\sin qr}{qr} \,dr\
\end{equation}

\noindent where $g(r)$ 
is calculated using a method based on the Ornstein-Zernike relation and the hybridized-mean spherical approximation (HMSA) closure relation first proposed by Zerah and Hansen~\cite{Zerah1986}.   

The apparent hydrodynamic radius $R_\mathrm{h,app}$ is calculated using the relationship between the short time collective diffusion coefficient $D_\mathrm{c}^\mathrm{s}(q)$ and the ideal diffusion coefficient $D_0$ in the absence of interactions, given by \cite{Naegele1996, Banchio2008}
 \begin{align}
	\label{Dcoll}
	D_\mathrm{c}^\mathrm{s}(q) = D_0 \frac{H(q)}{S(q)}
\end{align}
where $H(q)$ is the hydrodynamic function that describes the effects of hydrodynamic interactions. $S(q)$ is calculated using HMSA as described above. For the calculation of $H(q)$ we assume pairwise additive hydrodynamic interactions, which should be accurate up to volume fractions of around $\varphi \leq 0.05$. For our antibodies this roughly corresponds to $c \leq 25$ mg/ml. We can thus write~\cite{Neal1999}
\begin{align}
	\label{Hq}
	H(q) = 1 + 6 \pi \rho R_\mathrm{h}  \int_{0}^{\infty} r(g(r) - 1) F(qr) \,dr\ .
\end{align}
where $F(qr)$ is given by
\begin{align}
	\label{Fqr}
    F(qr) = \frac{\sin qr}{qr} +  \frac{\cos qr}{(qr)^2} -  \frac{\sin qr}{(qr)^3}.
    \end{align}

For small particles such as proteins, the measured diffusion coefficient corresponds to the so-called gradient diffusion coefficient given by 
\begin{align}
	\label{Dc}
	D_\mathrm{c} = \lim_{q \to 0} D_\mathrm{c}^\mathrm{s}(q) = D_0 \frac{H(0)}{S(0)} 
\end{align}
\noindent where $H(0)$ is related to the sedimentation velocity~\cite{Banchio2008}. This results in
\begin{align}
	\label{Rhapp}
	R_\mathrm{h,app} = R_{\mathrm{h},0} \frac{S(0)}{H(0)} 
\end{align}
\noindent where $R_{\mathrm{h},0}$ is the hydrodynamic radius of the antibodies in the absence of interactions, i.e., at infinite dilution.

\normalsize




\section*{Acknowledgements}

We thank Jeffrey Everts, Marco Polimeni and Letizia Tavagnacco for useful discussions, and Alessandro Gulotta and Szilard Sáringer for providing the experimental structure factors. F.C. and P.S. acknowledge funding from the European Union’s Horizon Europe research and innovation programme under the grant agreement number 101106720, Marie Skłodowska-Curie Action Postdoctoral Fellowship, project many(Anti)Bodies - mAB. S.M.-A. and E.Z. acknowledge funding from the European Union’s Horizon Europe research and innovation programme under the grant agreement number 101106848, Marie Skłodowska-Curie Action Postdoctoral Fellowship, project MGELS. A.S. acknowledges funding from the Swedish Research Council (VR; Grant No. 2022-03142). The computer simulations were also enabled by resources provided by the National Academic Infrastructure for Supercomputing in Sweden (NAISS) and the Swedish National Infrastructure for Computing (SNIC) at Lund University, partially funded by the Swedish Research Council through grant agreements no. 2022-06725 and no. 2018-05973. 
This work is also part of the “LINXS Antibodies in Solution” research program and we acknowledge the financial support by the LINXS Institute of Advanced Neutron and X-ray Science. 

\section*{Author Contributions}
Author contributions are defined based on CRediT (Contributor Roles Taxonomy). Conceptualization: F.C., P.S., E.Z.; Formal analysis: F.C., S.M.-A.; Funding acquisition: F.C., S.M.-A., A.S., P.S., E.Z.; Investigation: F.C., S.M.-A., A.S., P.S., E.Z.; Methodology: F.C., S.M.-A.; Project administration: F.C., P.S.; Software: F.C., S.M.-A.; Supervision: P.S., E.Z.; Validation: F.C., S.M.-A.; Visualization: F.C.; Writing – original draft: F.C.; Writing – review and editing: F.C., S.M.-A., A.S., P.S., E.Z..
			

\clearpage
\newpage
\onecolumngrid
\begin{center}
\large
\textbf{Deciphering Molecular Charge Anisotropy:\\the Case of Antibody Solutions\\ \bigskip Supplementary Information}

\normalsize
\bigskip
Fabrizio Camerin\textsuperscript{1}, Susana Marín-Aguilar\textsuperscript{2}, Anna Stradner\textsuperscript{1, 3},\\ Peter Schurtenberger\textsuperscript{1, 3}, Emanuela Zaccarelli\textsuperscript{4, 2}\\
\medskip
\small
\textit{%
\textsuperscript{1}Division of Physical Chemistry, Department of Chemistry, Lund University, Lund, Sweden\\
\textsuperscript{2}Department of Physics, Sapienza University of Rome, Piazzale Aldo Moro 2, 00185 Roma, Italy\\
\textsuperscript{3}LINXS Institute of Advanced Neutron and X-ray Science, Lund University, Lund, Sweden\\
\textsuperscript{4}CNR Institute of Complex Systems, Uos Sapienza, Piazzale Aldo Moro 2, 00185 Roma, Italy\\
}

\end{center}

\normalsize
\bigskip
\renewcommand{\theequation}{S\arabic{equation}}\setcounter{equation}{0}
\renewcommand{\thefigure}{S\arabic{figure}}\setcounter{figure}{0}
\renewcommand{\thetable}{S\arabic{table}}\setcounter{table}{0}

\section{Numbering coarse-grained model}

\begin{figure}[h!]
\centering
\includegraphics[width=0.75\textwidth]{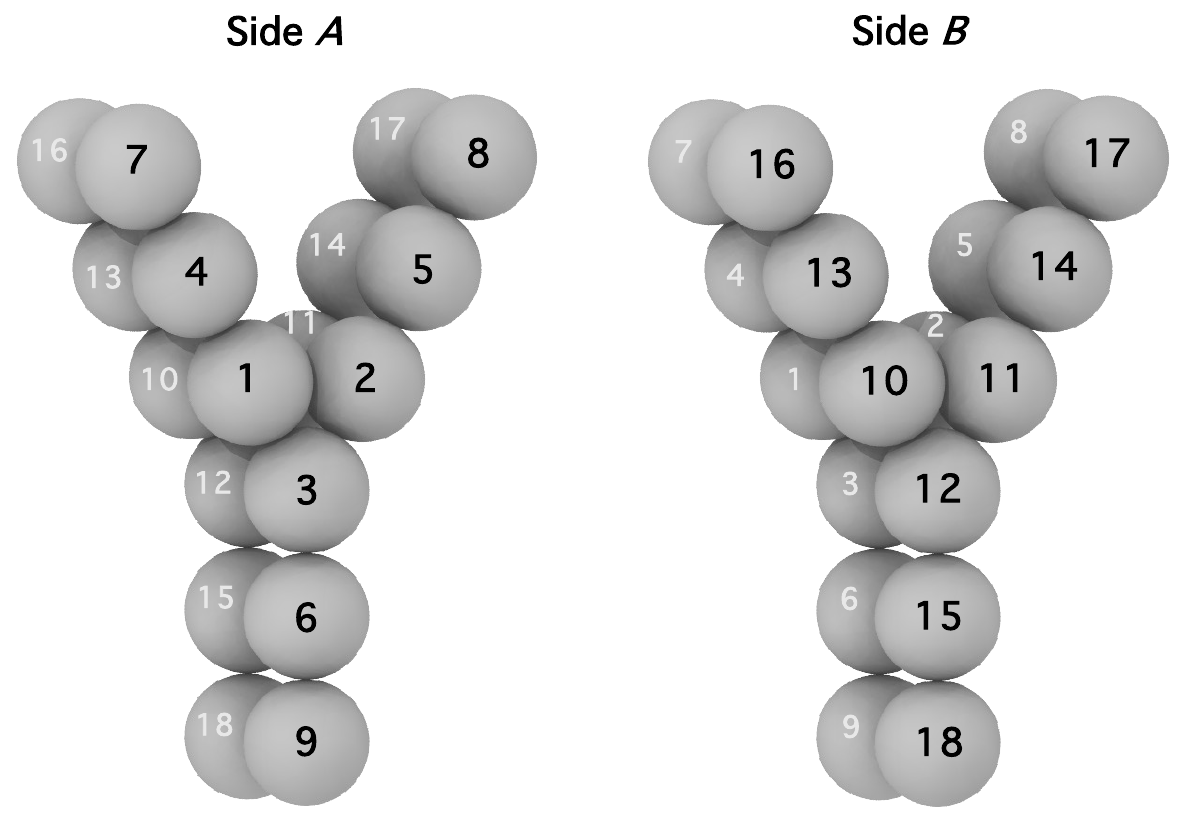}
\caption{\small \textbf{Numbering of the coarse-grained model.} Numbering of the coarse-grained 18-bead model for antibodies employed in this work for side $A$ and $B$ of the antibody as defined in the main text.}
\label{fig:numbering}
\end{figure}

\clearpage
\newpage

\section{Effect on varying parameters on charges}

\begin{figure}[h!]
\centering
\includegraphics[width=0.75\textwidth]{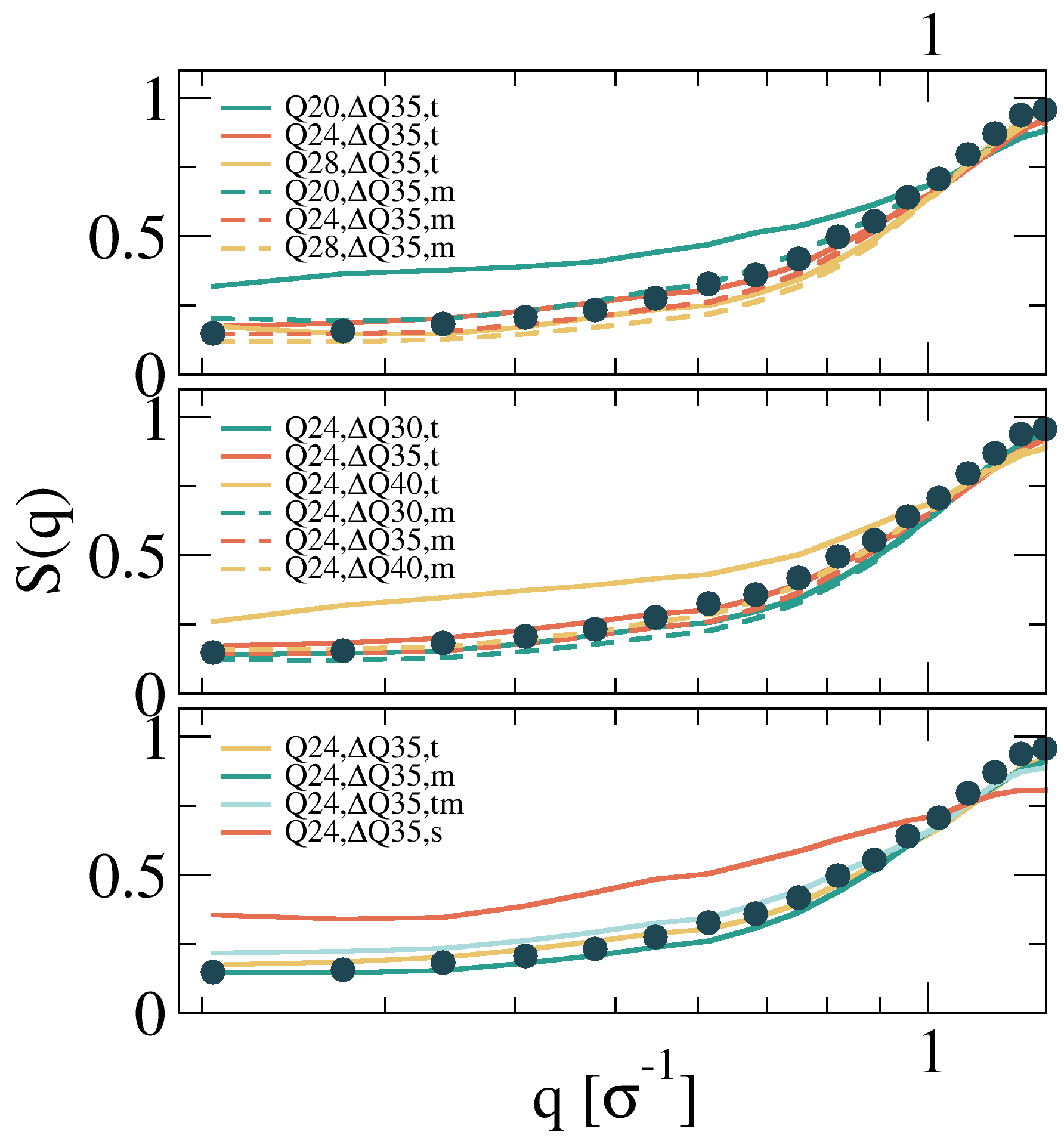}
\caption{\small \textbf{Varying parameters.} Static structure factor $S(q)$ as a function of the wavevectors $q$ for (a) varying the total charge $Q$, (b) varying the charge difference $\Delta Q$ and (c) the charge distribution $\mathcal{Q}$, keeping fixed the other parameters, for $c=100$ mg/ml. Indicated in the legends is the nominal value of the parameters, while the actual one may slightly change (see Results and Methods). As a reference, circles indicate the experimental $S(q)$ for $c=100$ mg/ml.}
\label{fig:models}
\end{figure}

\clearpage
\newpage





\section{Features of the initial training dataset}

\begin{table}[h]
\begin{adjustbox}{width=1\textwidth}
\begin{tabular}{r r r r r c r r r r r r r r r r r r r r r r r r}
\toprule
No. & $Q_{tot}^{tar}$ & $Q_{tot}^{act}$ & $\Delta Q^{tar}$ & $\Delta Q^{act}$ & Loc & 1 & 2 & 3 & 4 & 5 & 6 & 7 & 8 & 9 & 10 & 11 & 12 & 13 & 14 & 15 & 16 & 17 & 18 \\
\midrule
1 & 20 & 20 & 30 & 32 & m & +1 & +1 & +3 & +1 & +1 & +0 & +2 & +4 & +1 & +5 & +2 & +1 & -1 & -4 & -1 & +3 & +0 & +1 \\
2 & 20 & 12 & 30 & 30 & s & +1 & +1 & +3 & +1 & +1 & +1 & +4 & +6 & +3 & -1 & -2 & -1 & -1 & -1 & -2 & +0 & +0 & -1 \\
3 & 20 & 20 & 30 & 32 & tm & +1 & +1 & +3 & +1 & +1 & +0 & +2 & +4 & +1 & +6 & +5 & +1 & -1 & +0 & -3 & -1 & +0 & -1 \\
4 & 20 & 20 & 30 & 32 & t & +1 & +1 & +3 & +1 & +1 & +0 & +2 & +4 & +1 & +0 & +1 & +6 & +2 & +1 & +2 & -1 & -4 & -1 \\
5 & 20 & 17 & 35 & 35 & m & +1 & +1 & +3 & +1 & +1 & +0 & +2 & +4 & +1 & +1 & +0 & +1 & -3 & -2 & -4 & +6 & +1 & +3 \\
6 & 20 & 17 & 35 & 35 & s & +1 & +1 & +3 & +3 & +3 & +2 & +4 & +6 & +3 & -1 & -1 & -1 & -2 & -2 & +0 & +0 & -1 & -1 \\
7 & 20 & 20 & 35 & 32 & tm & +1 & +1 & +3 & +1 & +1 & +0 & +2 & +4 & +1 & +0 & +6 & +6 & -1 & -1 & -1 & +0 & -1 & -2 \\
8 & 20 & 17 & 35 & 35 & t & +1 & +1 & +3 & +1 & +1 & +0 & +2 & +4 & +1 & +3 & +5 & +0 & +1 & +0 & +3 & -4 & -2 & -3 \\
9 & 20 & 20 & 40 & 38 & m & +1 & +1 & +3 & +1 & +1 & +0 & +2 & +5 & +3 & +1 & +5 & +1 & -3 & -5 & -1 & +3 & +1 & +1 \\
10 & 20 & 20 & 40 & 38 & s & +1 & +2 & +5 & +3 & +3 & +2 & +4 & +6 & +3 & +0 & -1 & -2 & -2 & -1 & -1 & +0 & -1 & -1 \\
11 & 20 & 17 & 40 & 41 & tm & +1 & +1 & +3 & +1 & +1 & +0 & +2 & +4 & +1 & +6 & +6 & +3 & -3 & +0 & +0 & -5 & -1 & -3 \\
12 & 20 & 20 & 40 & 38 & t & +1 & +1 & +3 & +1 & +1 & +0 & +2 & +5 & +3 & +0 & +2 & +2 & +2 & +3 & +3 & -2 & -2 & -5 \\
13 & 24 & 24 & 30 & 30 & m & +1 & +1 & +3 & +1 & +1 & +0 & +2 & +4 & +2 & +2 & +1 & +2 & +0 & -1 & -2 & +3 & +2 & +2 \\
14 & 24 & 12 & 30 & 30 & s & +1 & +1 & +3 & +1 & +1 & +1 & +4 & +6 & +3 & -1 & +0 & -1 & -1 & -3 & +0 & -1 & -1 & -1 \\
15 & 24 & 20 & 30 & 32 & tm & +1 & +1 & +3 & +1 & +1 & +0 & +2 & +4 & +1 & +3 & +5 & +4 & -1 & -2 & -1 & -1 & -1 & +0 \\
16 & 24 & 24 & 30 & 30 & t & +1 & +1 & +3 & +1 & +1 & +0 & +2 & +4 & +2 & +2 & +4 & +1 & +2 & +2 & +1 & -3 & +0 & +0 \\
17 & 24 & 23 & 35 & 35 & m & +1 & +1 & +3 & +1 & +1 & +0 & +2 & +5 & +3 & +2 & +1 & +5 & -1 & -3 & -2 & +3 & +1 & +0 \\
18 & 24 & 17 & 35 & 35 & s & +1 & +1 & +3 & +3 & +3 & +2 & +4 & +6 & +3 & -1 & -2 & +0 & -1 & -1 & -1 & -1 & +0 & -2 \\
19 & 24 & 23 & 35 & 35 & tm & +1 & +1 & +3 & +1 & +1 & +0 & +2 & +4 & +1 & +3 & +6 & +6 & -1 & -1 & -1 & +0 & -1 & -2 \\
20 & 24 & 23 & 35 & 35 & t & +1 & +1 & +3 & +1 & +1 & +0 & +2 & +5 & +3 & +5 & +1 & +3 & +1 & +1 & +1 & -4 & +0 & -2 \\
21 & 24 & 23 & 40 & 41 & m & +1 & +1 & +3 & +1 & +1 & +0 & +2 & +4 & +1 & +1 & +4 & +4 & -4 & -5 & +0 & +3 & +5 & +1 \\
22 & 24 & 22 & 40 & 40 & s & +2 & +3 & +5 & +3 & +3 & +2 & +4 & +6 & +3 & -2 & -1 & -1 & -1 & +0 & +0 & -1 & -1 & -2 \\
23 & 24 & 26 & 40 & 38 & tm & +1 & +1 & +3 & +1 & +1 & +0 & +2 & +4 & +1 & +6 & +6 & +6 & -1 & -1 & -1 & -1 & -1 & -1 \\
24 & 24 & 23 & 40 & 41 & t & +1 & +1 & +3 & +1 & +1 & +0 & +2 & +4 & +1 & +2 & +2 & +4 & +3 & +2 & +5 & -4 & -2 & -3 \\
25 & 28 & 24 & 30 & 30 & m & +1 & +1 & +3 & +1 & +1 & +0 & +2 & +4 & +2 & +1 & +3 & +0 & -1 & -2 & +0 & +5 & +1 & +2 \\
26 & 28 & 12 & 30 & 30 & s & +1 & +1 & +3 & +1 & +1 & +1 & +4 & +6 & +3 & -1 & -1 & -2 & -1 & -1 & -1 & +0 & +0 & -2 \\
27 & 28 & 23 & 30 & 35 & tm & +1 & +1 & +3 & +1 & +1 & +0 & +2 & +4 & +1 & +4 & +6 & +5 & -1 & -1 & +0 & +0 & -2 & -2 \\
28 & 28 & 24 & 30 & 30 & t & +1 & +1 & +3 & +1 & +1 & +0 & +2 & +4 & +2 & +1 & +0 & +2 & +1 & +2 & +6 & -2 & -1 & +0 \\
29 & 28 & 29 & 35 & 35 & m & +1 & +1 & +3 & +1 & +1 & +0 & +2 & +4 & +1 & +6 & +2 & +4 & -1 & +0 & -2 & +4 & +0 & +2 \\
30 & 28 & 17 & 35 & 35 & s & +1 & +1 & +3 & +3 & +3 & +2 & +4 & +6 & +3 & -2 & -1 & +0 & -1 & -1 & +0 & -2 & -1 & -1 \\
31 & 28 & 23 & 35 & 35 & tm & +1 & +1 & +3 & +1 & +1 & +0 & +2 & +4 & +1 & +6 & +6 & +3 & -2 & -1 & -1 & -1 & +0 & -1 \\
32 & 28 & 29 & 35 & 35 & t & +1 & +1 & +3 & +1 & +1 & +0 & +2 & +4 & +1 & +1 & +2 & +5 & +1 & +3 & +6 & +0 & -2 & -1 \\
33 & 28 & 28 & 40 & 40 & m & +1 & +1 & +3 & +1 & +1 & +0 & +2 & +4 & +3 & +2 & +1 & +6 & -4 & -1 & -1 & +1 & +3 & +5 \\
34 & 28 & 22 & 40 & 40 & s & +2 & +3 & +5 & +3 & +3 & +2 & +4 & +6 & +3 & -1 & -1 & +0 & -1 & -1 & -1 & -2 & -1 & -1 \\
35 & 28 & 28 & 40 & 40 & tm & +1 & +1 & +3 & +1 & +1 & +0 & +2 & +4 & +3 & +6 & +6 & +6 & -1 & -1 & -1 & -1 & -1 & -1 \\
36 & 28 & 28 & 40 & 40 & t & +1 & +1 & +3 & +1 & +1 & +0 & +2 & +4 & +3 & +2 & +5 & +1 & +4 & +4 & +2 & -3 & -2 & -1 \\
\bottomrule
\end{tabular}
\end{adjustbox}
\caption{\small \textbf{Initial training dataset.} Features of the first training dataset reporting: the ID No., the target total charge $Q_{tot}^{tar}$, the actual total charge $Q_{tot}^{act}$, the target difference in charge $\Delta Q_{tot}^{tar}$, the actual difference in charge $\Delta Q_{tot}^{act}$, the position where the negative charges are distributed (tips $t$, middle $m$, tips and middle $tm$, spread $s$), and the charge for each of the $18$ beads, according to the classification given in Figure~\ref{fig:numbering}.}
\label{fig:models}
\end{table}

\begin{figure}[h!]
\centering
\includegraphics[width=\textwidth]{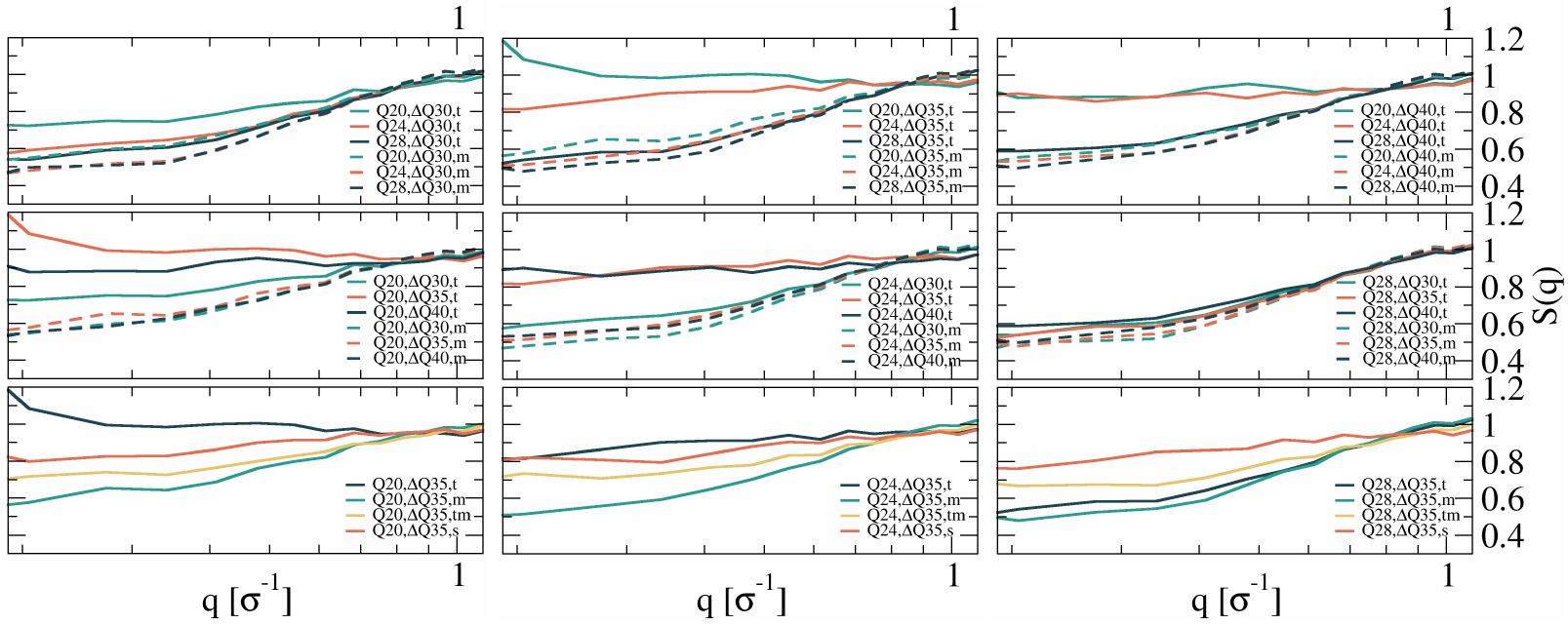}
\caption{\small \textbf{Static properties of the initial training dataset.} Static structure factors $S(q)$ for different combinations of parameters used as training dataset of the neural network in the first optimization cycle.}
\label{fig:models}
\end{figure}

\begin{figure}[h!]
\centering
\includegraphics[width=\textwidth]{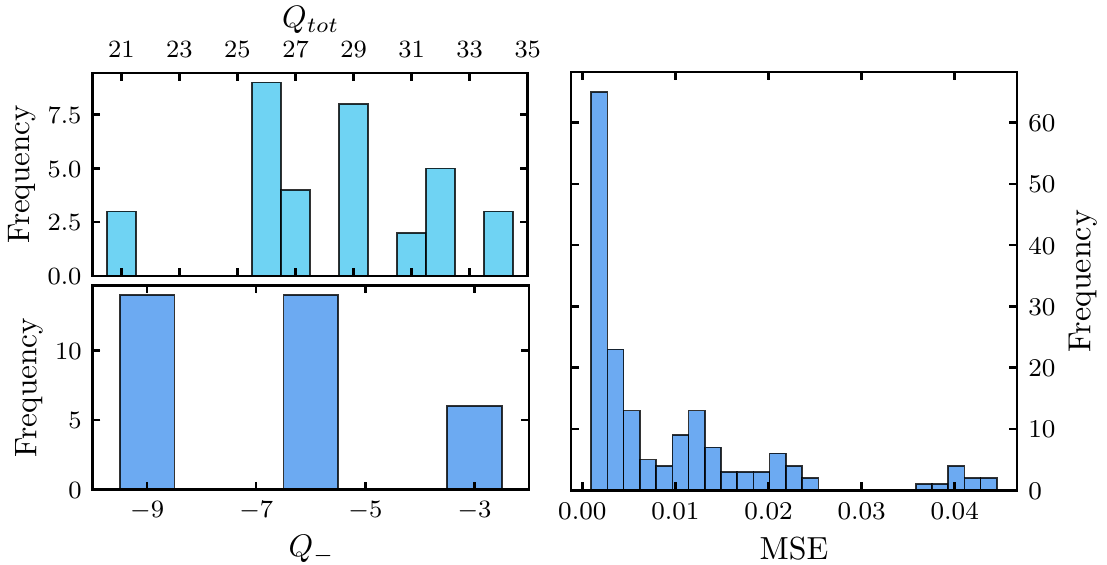}
\caption{\small \textbf{Additional features for the initial training dataset.} Distribution of (a) total charge, (b) sum of the negative charges, and (c) mean squared error (MSE) for the antibody configurations used as dataset for the training in the first optimization cycle.}
\label{fig:models}
\end{figure}

\clearpage
\newpage

\section{Complete list of MSE for each optimization cycle}


\begin{table}[h]
\renewcommand{\arraystretch}{1.5}
\begin{adjustbox}{width=1\textwidth}
\begin{tabular}{|c|c| c c c c c c c c c c c c c c c c c c |c|c|c|}
\hline
\multirow{2}{*}{\parbox{2.2cm}{\centering optimization cycle}} & \multirow{2}{*}{ranking} & \multicolumn{18}{|c|}{charges for each bead} & \multicolumn{3}{|c|}{MSE} \\ \cline{3-23}
 &  & 1 & 2&3&4&5&6&7&8&9&10&11&12&13&14&15&16&17&18 & $c=20$ mg/ml & $c=100$ mg/ml & Mean \\\hline
\multirow{3}{*}{I} & 1 & +1 &+1& +3& +1& +1& +0 &+2 &+4 &+1 &+1& +4& +5& +0& +1& +1& -1& -2& -2 &  0.0024 & 0.0005 & 0.0014 \\ \cline{2-2} \cline{21-23}
& 2 & +1 & +1 &+3 &+1 &+1 &+0 &+2& +4& +1 &+1 &+2 &+4 &+1 &+2& +4 &-1 &-2 &-1  & 0.0012 & 0.0028 & 0.0019\\  \cline{2-2} \cline{21-23}
& 3 & +1 &+1 &+3& +1 &+1& +0 &+2& +4 &+1& +1 &+1& +4& +1& +2 &+5& -1& -1 &-1  & 0.0017 & 0.0029 & 0.0023\\  \cline{2-2} \cline{21-23}
\hline
\multirow{3}{*}{II} & 1 & +1 &+1& +3& +1& +1& +0& +2& +4& +1& +3& +1& +2& +2& +1& +4& -2& +0& -3 & 0.0011 & 0.0004  & 0.0008\\  \cline{2-2} \cline{21-23}
& 2 & +1& +1& +3& +1& +1& +0& +2& +4& +1& +1& +1& +2& +2& +3& +3& +0& -1& -2  & 0.0013 & 0.0030 & 0.0022\\ \cline{2-2} \cline{21-23}
& 3 & +1 &+1 &+3 &+1 &+1 &+0 &+2 &+4 &+1 &+2 &+1 &+2 &+3 &+1 &+5 &-1 &+0 &-2  & 0.0016 & 0.0031 & 0.0023\\  \hline
\multirow{3}{*}{III} & 1 & +1& +1 &+3 &+1 &+1 &+0 &+2 &+4 &+1 &+2 &+2 &+1 &+3 &+1 &+4 &-1 &+0 &-3& 0.0006 & 0.0009  & 0.00076 \\  \cline{2-2} \cline{21-23}
& 2 & +1& +1 &+3 &+1& +1& +0& +2& +4& +1& +3& +1& +2& +2& +1& +4& -1& +0& -3  & 0.0007 & 0.0008& 0.00077 \\ \cline{2-2} \cline{21-23}
& 3 & +1 &+1 &+3& +1& +1& +0& +2& +4& +1& +2& +1& +2& +3& +1 &+5 &-1 &+0 &-3  & 0.0006 & 0.0010 & 0.00082 \\  \cline{2-2} \cline{21-23}
\hline
\end{tabular}
\end{adjustbox}
\caption{\small \textbf{Output details of the optimization process.} Mean squared error (MSE) for the outputs of the neural network for $c=20$ mg/ml and $c=100$ mg/ml. These are reported in order from the lowest to the highest optimization cycle with the complete list of charges for each bead, from A1 to B18, according to the classification given in Figure~\ref{fig:numbering}.}
\label{tab:msecyles}
\end{table}

\section{Static structure factors for the best combinations overall}

\begin{figure}[h!]
\centering
\includegraphics[width=0.75\textwidth]{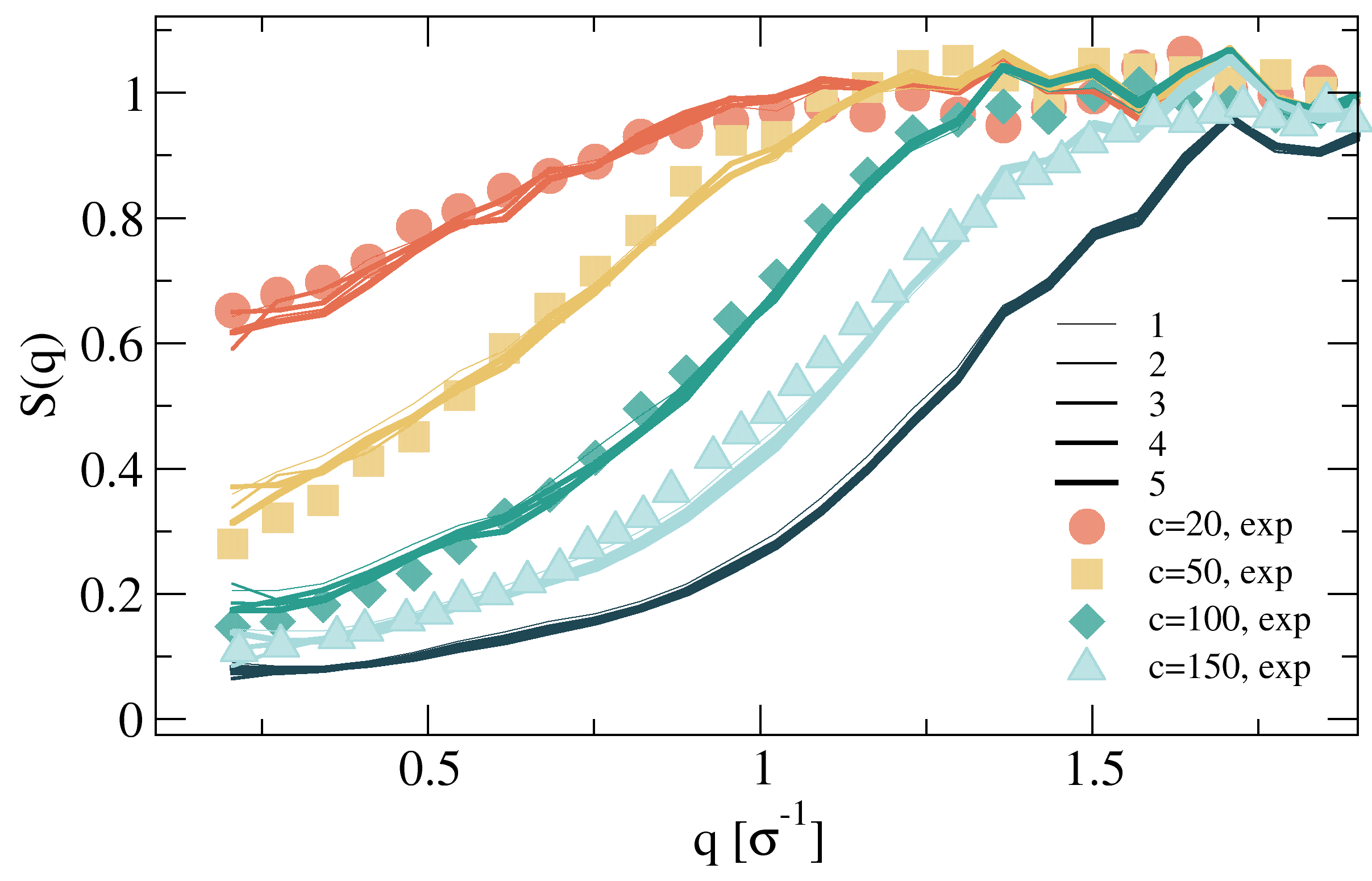}
\caption{\small \textbf{Best combinations overall.} Static structure factors for the five configurations identified as "best overall" in the main text for different concentrations $c=20,50,100,150,200$ mg/ml, as compared to experiments. Experimental data for $c=200$ mg/ml are not available.}
\label{fig:models}
\end{figure}

\clearpage
\newpage

\section{Complete list of features used for the SHAP algorithm}

\begin{table}[h]
\renewcommand{\arraystretch}{1.5}
\begin{tabular}{|c|c|c|c|}
\hline
Ranking & Feature name & Feature details & SHAP value \\ \hline
1 & $\mu_x$  & $\sum_{i=1}^{18} \tilde{q}_{i} r_{xi}$& 0.0087 \\ \hline
2 & $Q_-$  &$\sum_i \tilde{q}_i<0$ & 0.0083 \\ \hline
3 & $Q_{tm}$  & $Q_t+Q_m$& 0.0041 \\ \hline
4 & $Q$  &$\sum_i \tilde{q}_i$ & 0.0031 \\ \hline
5 & $Q_t$  & $\tilde{q}_{16}+\tilde{q}_{17}+\tilde{q}_{18}$& 0.0030 \\ \hline
6 & $Q_{side\,A}$  & $\sum_{i=1}^{9} \tilde{q}_i$& 0.0029 \\ \hline
7 & $Q_{side\,B}$  & $\sum_{i=10}^{18} \tilde{q}_i$& 0.0023 \\ \hline
8 & $|\mu|$  & $\sqrt{\mu_x^2+\mu_y^2+\mu_z^2}$ & 0.0020 \\ \hline
9 & $Q_{center}$  &  $\tilde{q}_{10}+\tilde{q}_{11}+\tilde{q}_{12}$& 0.0019 \\ \hline
10 & $\mu_y$  & $\sum_{i=1}^{18} \tilde{q}_{i} r_{yi}$& 0.0018 \\ \hline
11 & prod5  &$\tilde{q}_{15}\times \tilde{q}_{18}$ & 0.00078 \\ \hline
12 & $\mu_z$  &$\sum_{i=1}^{18} \tilde{q}_{i} r_{zi}$ & 0.00075 \\ \hline
13 & prod6  & $\tilde{q}_{13}\times \tilde{q}_{16}$& 0.00066 \\ \hline
14 & prod1  &$\tilde{q}_{14}\times \tilde{q}_{17}$ & 0.00058 \\ \hline
15 & prod4  & $\tilde{q}_{12}\times \tilde{q}_{15}$& 0.00056 \\ \hline
16 & $\Delta Q$  &$Q_+-Q_-$ & 0.00052 \\ \hline 
17 & $Q_m$  & $\tilde{q}_{13}+\tilde{q}_{14}+\tilde{q}_{15}$& 0.00051 \\ \hline 
18 & prod8  &$\tilde{q}_{10}\times \tilde{q}_{11}$ & 0.00047 \\ \hline
19 & prod2  & $\tilde{q}_{11}\times \tilde{q}_{14}$& 0.00045 \\ \hline
20 & prod7  & $\tilde{q}_{10}\times \tilde{q}_{13}$& 0.00045 \\ \hline
21 & prod3  &$\tilde{q}_{11}\times \tilde{q}_{12}$ & 0.00042 \\ \hline
22 & prod9  & $\tilde{q}_{10}\times \tilde{q}_{11}$& 0.00040 \\ \hline
23 & $Q_+$  &$\sum_i \tilde{q}_i>0$ & 0.00029 \\ \hline
\end{tabular}
\caption{\small \textbf{SHAP features} Features used as input for the SHAP algorithm as described in Methods and respective SHAP value. These are given in order from the most to the least important in determining the static structure factors of the antibodies.}
\label{tab:msecyles}
\end{table}


\twocolumngrid
\clearpage
\newpage

\bibliography{bibliography.bib}

\end{document}